\documentclass{ieeeaccess}
\usepackage{cite}
\usepackage{amsmath,amssymb,amsfonts}
\usepackage{algorithm}
\usepackage{graphicx}
\usepackage{textcomp}
\usepackage{enumitem,dsfont,algpseudocode}
\def\BibTeX{{\rm B\kern-.05em{\sc i\kern-.025em b}\kern-.08em
    T\kern-.1667em\lower.7ex\hbox{E}\kern-.125emX}}

\usepackage{acronym}

\definecolor{NewColor}{rgb}{0.2,0,0.5}

\newcommand{\myVec}[1]{{{#1}}}
\newcommand{\myMat}[1]{{{#1}}}
\newcommand{\mySet}[1]{\mathcal{#1}}

\newcommand{\InputSpace}{\mySet{X}}

\newcommand{\Input}{\myVec{x}}
\newcommand{\Label}{\myVec{s}^{\rm true}}
\newcommand{\Output}{\myVec{s}}
\newcommand{\LabelSpace}{\mySet{S}}

\newcommand{\Ntraining}{n_t}

\newcommand{\PdfNew}[1]{p}

\newcommand{\E}{\mathds{E}}	

\newcommand{\ObjParam}{\myVec{\theta}^{\rm o}}
\newcommand{\HypParam}{\myVec{\theta}^{\rm h}}

\newcommand{\csMatrix}{\myMat{H}}

\newcommand{\Distribution}{\mySet{P}}
\newcommand{\Data}{\mySet{D}}

\newtheorem{example}{Example}

\acrodef{cnn}[CNN]{convolutional neural network} 
\acrodef{relu}[ReLU]{rectified linear unit}
\acrodef{dnn}[DNN]{deep neural network}
\acrodef{adc}[ADC]{analog-to-digital convertor}
\acrodef{ista}[ISTA]{iterative soft thresholding algorithm}
\acrodef{lista}[LISTA]{learned \ac{ista}}
\acrodef{cs}[CS]{compressed sensing}
\acrodef{bp}[BP]{belief propagation}
\acrodef{bpsk}[BPSK]{binary phase shift keying}
\acrodef{dtft}[DTFT]{discrete-time Fourier transform}
\acrodef{dnn}[DNN]{Deep neural network} 
\acrodef{gan}[GAN]{generative adversarial network} 
\acrodef{gnn}[GNN]{graph neural network} 
\acrodef{gru}[GRU]{gated recurrent unit} 
\acrodef{lstm}[LSTM]{long short-term memory} 
\acrodef{csi}[CSI]{channel state information}
\acrodef{map}[MAP]{maximum a-posteriori probability}
\acrodef{snr}[SNR]{signal-to-noise ratio}
\acrodef{bs}[BS]{base station} 
\acrodef{em}[EM]{expectation maximization} 
\acrodef{iot}[IOT]{Interent of Things}
\acrodef{mimo}[MIMO]{multiple-input multiple-output}
\acrodef{mse}[MSE]{mean-squared error}
\acrodef{pdf}[PDF]{probability density function}
\acrodef{rv}[RV]{random variable}
\acrodef{fec}[FEC]{forward error correction} 
\acrodef{lti}[LTI]{linear time-invariant}
\acrodef{wss}[WSS]{wide-sense stationary}
\acrodef{psd}[PSD]{power spectral density}
\acrodef{ser}[SER]{symbol error rate} 
\acrodef{ber}[BER]{bit error rate} 
\acrodef{sgd}[SGD]{stochastic gradient descent}  
\acrodef{awgn}[AWGN]{additive white Gaussian noise} 
\acrodef{ut}[UT]{user terminal}  
\acrodef{ml}[ML]{machine learning}  
\acrodef{rnn}[RNN]{recurrent neural network} 
\acrodef{fc}[FC]{fully-connected}
\acrodef{sic}[SIC]{soft interference cancellation}
\acrodef{pmf}[PMF]{probability mass function}
\acrodef{sp}[SP]{sum-product} 
\acrodef{ista}[ISTA]{iterative soft thresholding algorithm}
\acrodef{pca}[PCA]{principal component analysis}
\acrodef{svd}[SVD]{singular value decomposition}
\acrodef{gan}[GAN]{generative adversarial network}
\acrodef{admm}[ADMM]{alternating direction method of multipliers}
\acrodef{lqg}[LQG]{linear-quadratic-Gaussian}
\acrodef{bptt}[BPTT]{backpropagation through time}

\begin{document}
\history{Date of publication xxxx 00, 0000, date of current version xxxx 00, 0000.}
\doi{10.1109/ACCESS.2017.DOI}

\title{Model-Based Deep Learning: On the Intersection of Deep Learning and Optimization}
\author{\uppercase{Nir Shlezinger}\authorrefmark{1}, \IEEEmembership{Member, IEEE},
\uppercase{Yonina C. Eldar}\authorrefmark{2}, \IEEEmembership{Fellow, IEEE}, and \uppercase{Stephen P. Boyd}\authorrefmark{3},
\IEEEmembership{Fellow, IEEE}}
\address[1]{School of Electrical and Computer Engineering, Ben-Gurion University of the Negev, Beer-Sheva 8410501, Israel (e-mail: nirshl@bgu.ac.il)}
\address[2]{Faculty of Mathematics and Computer Science, Weizmann Institute of Science, Rehovot 7610001, Israel (e-mail: yonina.eldar@weizmann.ac.il)}
\address[3]{Department of Electrical Engineering, Stanford University, Palo Alto, CA, USA (e-mail: boyd@stanford.edu)}
\tfootnote{The work was supported in part by ACCESS (AI Chip Center for Emerging Smart Systems), Stanford SystemX, by the Igel Manya Center for Biomedical
Engineering and Signal Processing, and by the European Research Council (ERC) under 
the European Union’s Horizon 2020 research and innovation programme (grant agreement No. 101000967).}

\markboth
{Shlezinger \headeretal: Model-Based Deep Learning: On the Intersection of Deep Learning and Optimization}
{Shlezinger \headeretal: Model-Based Deep Learning: On the Intersection of Deep Learning and Optimization}

\corresp{Corresponding author: Nir Shlezinger (e-mail: nirshl@bgu.ac.il).}
.

\begin{abstract}
Decision making algorithms are used in a multitude of different applications. Conventional approaches for designing decision algorithms employ principled and simplified modelling, based on which one can determine decisions via tractable optimization. More recently,  deep learning approaches that use highly  parametric architectures tuned from data without relying on mathematical models, are becoming increasingly popular. Model-based optimization and data-centric deep learning are often considered to be distinct disciplines. Here, we characterize them as edges of a continuous spectrum varying in specificity and parameterization, and provide a tutorial-style presentation to the methodologies lying in the middle ground of this spectrum, referred to as {\em model-based deep learning}. We accompany our presentation with  running examples in super-resolution and stochastic control, and show how they are expressed using the provided characterization and specialized in each of the detailed methodologies. The gains of combining model-based optimization and deep learning are demonstrated using experimental results in various applications, ranging from biomedical imaging to digital communications. 
\end{abstract}


\titlepgskip=-15pt

\maketitle

\section{Introduction}
\label{sec:introduction}
\PARstart{O}{ptimization}  provides a framework for solving problems described in a tractable mathematical manner. Optimization-based methods have been successfully applied across a broad range of applications involving decision making, ranging from electrical engineering to control and finance. The conventional approach to carry out decision making involves the introduction of mathematical models for the problem and the solver based on domain knowledge. Such {\em model-based methods} form the basis for many classical and fundamental optimization techniques. Many of these classical approaches rely on simplified descriptions of the problem that make decision making tractable, computationally feasible, and interpretable. While model-methods  often work well, their simplified approximations can limit performance in some applications.

The unprecedented success of \ac{ml}, and particularly of deep learning, in areas such as computer vision and natural language processing \cite{lecun2015deep} gave rise  to methodology geared towards data. It is becoming common practice to replace principled task-specific decision mappings with abstract purely data-driven pipelines, trained with massive data sets. \acp{dnn} are trained end-to-end, often in a supervised manner, without relying on analytical approximations, and therefore, they can operate in scenarios where analytical models are unknown or highly complex \cite{Bengio09learning}. 
However, the abstractness and extreme parameterization of \acp{dnn} results in them  often being  treated as black-boxes; understanding how their predictions are obtained and how reliable they are tends to be quite challenging, and thus deep learning lacks the interpretability, flexibility, versatility, and reliability of model-based techniques.  

Due to the  limitations of model-based methods and data-driven pipelines, recent years have witnessed growing interest in decision mappings involving both principled mathematical optimization and data-centric deep learning \cite{shlezinger2020model,chen2021learning,maier2022known}. These include  frameworks such as deep unfolding \cite{monga2021algorithm} and learned optimization \cite{agrawal2021learning}, as well as task-specific techniques augmenting  optimizers with \acp{dnn} \cite{ahmad2020plug,bora2017compressed,satorras2019combining,shlezinger2021model}. While hybrid model-based deep learning methods are often designed and studied for specific tasks, their underlying methodology is  relevant to a broad range of applications, motivating the systematic characterization of the interplay between existing approaches. 

In this article we introduce a general framework  for model-based deep learning schemes. While classic optimization and deep learning are typically considered to be distinct disciplines, we  view them as edges of a continuum varying in specificity and parameterization. We build upon this characterization to provide a tutorial-style presentation of the main methodologies which lie in the middle ground of this spectrum, and combine model-based optimization with \ac{ml} as a form of model-based deep learning. Our  presentation is  exemplified with  running examples from super-resolution imaging and stochastic control.

We begin by providing a unified characterization for decision making algorithms, focusing on the main pillars of their design, which we identify as the decision rule type, the decision rule objective, and the evaluation procedure. Then, we show how classical model-based optimization as well as data-centric deep learning are obtained as special instances of this  unified characterization. We identify the components dictating the distinction between the methodologies in the formulated objectives, the corresponding decision rule types, and their associated parameters. We next present a spectrum of decision making approaches which vary in specificity and parameterization, with model-based optimization and deep learning constituting its edges, and provide a systematic categorization of model-based deep learning techniques into concrete strategies positioned along this continuous spectrum. The proposed categorization captures the interplay between the different techniques and their pros and cons in comparison with both model-based optimization and conventional deep learning. We present extensive experimental results applying model-based deep learning methodologies in various application areas, including ultrasound image processing, microscopy imaging, digital communications, and tracking of dynamic systems. The results   demonstrate the gains in performance and run-time of combining model-based optimization with deep learning over favouring  one discipline over the other.


\section{Decision Making}
	We consider a generic setup where the goal is to design a decision rule $f:\InputSpace\mapsto\LabelSpace$. The decision rule maps the context $\Input\in\InputSpace$, i.e., the available observations, into a decision  $\Output\in \LabelSpace$.

	{\bf Examples:}
	This generic formulation encompasses a multitude of settings involving classification, prediction, control, and many more. It thus corresponds to a broad range of  applications. The  task dictates the context space $\InputSpace$ and the possible decisions $\LabelSpace$.  A  partial list of such applications includes:
	\begin{itemize}
	    \item Signal processing -  The context $\Input$  includes samples from an observed signal or an image, which is mapped by $f$ into another signal (e.g., for denoising) or into some form of inference (e.g., anomaly detection). 
        \item Communications - The decision rule represents the operation of a digital receiver, which  decodes the channel output $\Input$ into an estimate of the transmitted message $\Output$.
	    \item  Vehicular control -   The decision rule $f$ is the control algorithm.   The context $\Input$ can include the traditional state variables, i.e., the vehicle sensory data, and commands.  The decision $\Output$ is the control action. 
        \item Finance -  The decision rule is the trading algorithm.  The context $\Input$ includes quantities such as financial forecasts and current positions.  The decision $\Output$ is the trade list, i.e., the list of assets to buy and to sell.   
	\end{itemize}

	\begin{figure}
	    \centering
	    \includegraphics[width=\columnwidth]{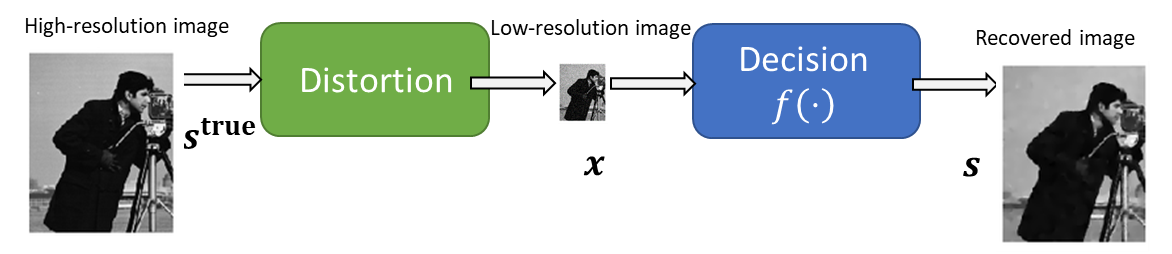}
	    \caption{Super-resolution recovery illustration.}
	    \vspace{-0.2cm}
	    \label{fig:ImageCompress1}
	\end{figure}

	\begin{figure}
	    \centering
	    \includegraphics[width=0.7\columnwidth]{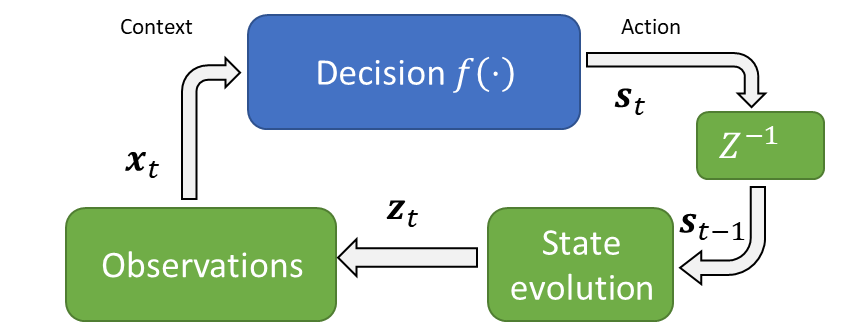}
	    \caption{Stochastic control illustration.}
	    \vspace{-0.2cm}
	    \label{fig:Control1}
	\end{figure}

	To keep the presentation focused, we repeatedly use two concrete running examples:
	\begin{example}[Super-Resolution]
	\label{exm:Inverse}
	Here, $\Output$ is a high-resolution image,  while $\Input$ is a distorted low-resolution version of the image. Thus, $\InputSpace$ and $\LabelSpace$ are the spaces of low-resolution and high-resolution images, respectively. Such decision rules, typically referred to as {\em recovery methods}, aim at reconstructing $\Label$ from its distorted version $\Input$, as illustrated in Fig.~\ref{fig:ImageCompress1}.
	\end{example}

	\begin{example}[Stochastic Control]
	\label{exm:Control}
    In our second  example, we consider a dynamic system, where the decision rule is a {\em control policy}. At each time period $t$, the goal is to map the noisy state observations $\Input_t$, where $\InputSpace$ is the space of possible sensory measurements, into an action $\Output_t$ within an action space $\LabelSpace$. The system is characterized by a latent state vector $\myVec{z}_t$ that evolves in a random fashion which is related to the previous state $\myVec{z}_{t-1}$ and action $\Output_{t-1}$, while being partially observable via the noisy $\Input_t$. This setup is illustrated in Fig.~\ref{fig:Control1}. 
	\end{example}

	\subsection{Decision Rule Types} 
	The above generic formulation allows the decision rule $f$ to be any mapping from $\InputSpace$ into $\LabelSpace$. In practice, decision rules are often carried out using a structured form. Some common types of decision rules are:
    \begin{enumerate}[label={\em T\arabic*}]
	    \item \label{itm:affine} An {\em affine  rule}, i.e., $\Output = \myMat{W} \Input +\myVec{b}$ for some $(\myMat{W}, \myVec{b})$. 
        \item \label{itm:tree} A {\em decision tree} chooses $\Output$ from a finite set of possible decisions $\{\Output_k\}$ by examining a set of nested conditions $\{{\rm cond}_k\}$, e.g., if ${\rm cond}_1(\Input)$ then $\Output = \Output_1$;  else inspect ${\rm cond}_2(\Input)$, and so on.  
        \item \label{itm:opt} An {\em optimization-based} decision rule chooses $\Output$ as a solution or approximate solution of an optimization problem parametrized by the context~$\Input$, i.e., ${\arg \min}_{\Output\in\LabelSpace}\mySet{L}(\Output;\Input)$, where $\mySet{L}$ is an objective function.
	    \item \label{itm:iter} An {\em iterative algorithm} finds its decision by executing a sequence of  mappings $h_k: \LabelSpace\times\InputSpace \mapsto \LabelSpace$, repeating $\Output_{k+1} = h_k\big(\Output_{k}; \Input\big)$ from an initial guess $\Output_{0}$ until convergence, or a fixed number of steps $K$, i.e., $\Output= h_{K}\big(h_{K-1}\big(\cdots h_1(\Output_{0};\Input);\Input\big);\Input\big)$.  
        \item \label{itm:dnn} A {\em neural network} is a special case of an iterative algorithm, where $h_k(\myVec{z}) = \sigma\big(\myMat{W}_k\myVec{z}+\myVec{b}_k\big)$ with $\sigma(\cdot)$ being an activation function and $(\myMat{W}_k,\myVec{b}_k)$ are parameters of the affine transformation.
        These mappings are referred to as layers. We have $\Output = h_{K}\big(h_{K-1}\big(\cdots h_1(\Input)\big)\big)$.
	\end{enumerate}
	The boundaries between decision rule types are not always clear, and there is some overlap between the  categories. For instance, an optimization-based decision rule with quadratic objective, where the context affects only the linear term in the objective, can be explicitly expressed as an affine decision rule.  As another example, iterative decision rules often arise as iterations that solve an optimization problem. Moreover, an iterative algorithm with $K$ iterations can often be viewed as a neural network, as we further elaborate on in the sequel. 
	
	Each of these decision rule types include {\em parameters}.  For example in an affine decision rule, the parameters are $\myMat{W}$ and $\myVec{b}$; in a decision tree, it is the values $\myVec{s}_k$ and parameters that specify the conditions.  In an iterative algorithm the parameters are those appearing in the functions $h_k$; and in a neural network, the parameters are  $\myMat{W}_k$ and $\myVec{b}_k$.
	In some cases the  number of parameters is small, such as decision trees with a small number of conditions. In other cases, e.g., when $f$ is a \ac{dnn}, decision rules can involve a massive number of parameters. These parameters capture the different mappings one can represent as decision rules.
	
	For a decision rule type, we let $\mySet{F}$ denote the set of possible
	decision rules, over all choices of parameters.
	In general, the more parameters there are, the broader the family of mappings captured by $\mySet{F}$, which in turn results in the decision rule capable of accommodating more diverse and generic functions.
Decision rules with fewer parameters are typically more specific,  capturing a limited family of mappings. 
Let ${\Theta}$  denote the parameter space for a decision rule family $\mySet{F}$, such that for each $\myVec{\theta}\in\Theta$, $f(\cdot; \myVec{\theta})$ is a mapping in $\mySet{F}$. 
We refer to the choosing of the decision rule parameters $ \myVec{\theta}$ as {\em tuning}. In principle, tuning can be carried out based on understanding and modeling of the task. In practice, tuning  typically involves simulation with either synthetic or real data; this procedure can be done manually when there are a few parameters, or by an automated algorithm for decision rules involving many parameters. In the latter case, tuning is also referred to as {\em training} or {\em learning}.

\subsection{Evaluating a Decision Rule}
The evaluation of a decision rule is comprised of two ingredients: $1)$ simulations where the decision rule is to be applied; and $2)$ an objective function for measuring its performance during the conducted simulations. 

\begin{figure}
    \centering
    \includegraphics[width=\columnwidth]{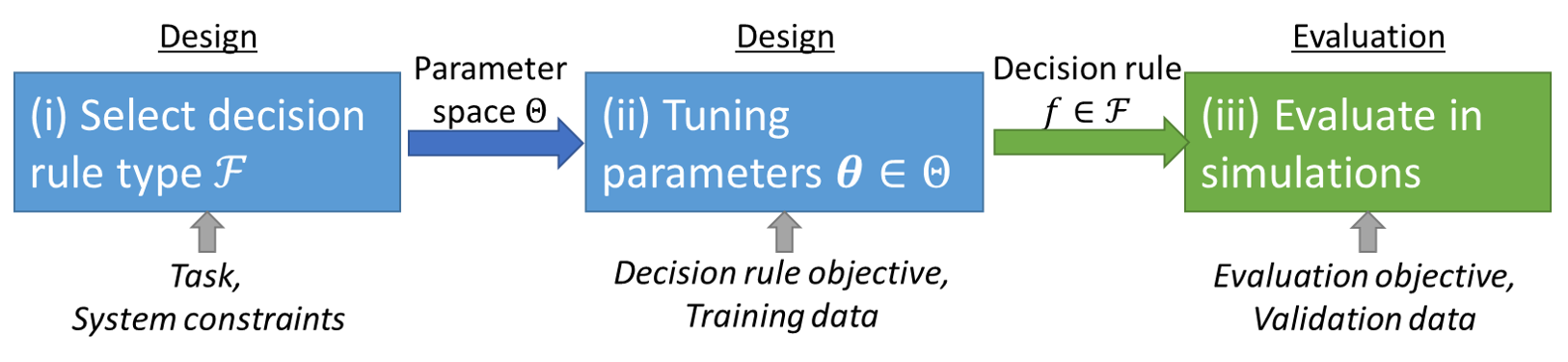}
    \caption{Decision rule selection procedure illustration.}
    \label{fig:DesignProc1}
    \vspace{-0.2cm}
\end{figure}

{\bf Simulations}  represent the setting in which the decision rule is required to operate after its parameters are tuned. They can be as simple as applying $f$ on held out data,  or involve complex mechanisms for emulating an overall system where the decision rule is to be applied and the expected environment. Various terminologies are used to describe simulators in different domains, including validation (in \ac{ml}), closed-loop simulation (in control), and back-testing (in finance).

\begin{example}
\label{exm:SRsim}
A simulation setup for super-resolution recovery (Example~\ref{exm:Inverse}) can be comprised of a set of unseen images and a mapping that converts them into low-resolution  data. 
\end{example}

\begin{example}
\label{exm:ControlSim}
A simulation setup for stochastic control (Example~\ref{exm:Control}) can be software which emulates how an action $\Output_{t-1}$ is translated into a state $\myVec{z}_t$ and an observed context $\myVec{x}_t$. 
\end{example}

{\bf Objective functions} are measures used to evaluate a decision rule.  
  In some cases, the objective is given by a cost or a loss function which one aims at minimizing, or it can be specified by an application utility or reward, which we wish to maximize.  In many applications there are multiple objectives, which  are scalarized into a single cost function, for example, by forming a weighted sum.   The objective  can be the average value of individual decisions, or a function of multiple decisions, e.g., a trajectory. In its most basic form, a loss function evaluates the decision rule for a given context as compared with some desired decision;  such a loss function is formulated as a mapping \cite{shalev2014understanding}
	  \begin{equation}
        \label{eqn:lossfunction}
        l:\mySet{F}\times\InputSpace\times\LabelSpace \mapsto \mySet{R}^+.
    \end{equation}
    Broadly speaking, \eqref{eqn:lossfunction} dictates the success criteria of a decision mapping for a given context-decision pair. 
    For instance, in inference tasks, candidate losses include  the error rate (zero-one) loss $        l_{\rm Err}(f,\Input,\Label) =  {\boldsymbol 1}_{f(\Input) \neq \Label}$ (for classification) and the $\ell_2$ loss $ l_{\rm Est}(f,\Input,\Label) = \|\Label - f(\Input)\|_2^2$ (for estimation).

    
    In optimization-based decision rules, the objective used to formulate the optimization problem  need not be the same as the evaluation objective function. 
    The {\em evaluation objective} measures the performance of the decision rule in simulation; it may be complex and capture multiple utilities of the overall system. In some applications, e.g., medical imaging, it may involve inspecting the simulations outcome by human experts. 
    The objective of the optimization-based decision rule, referred to as the {\em decision rule objective}, is used for tuning the decision rule; it is often a surrogate of the evaluation objective, including e.g., simplifications, approximations, and regularizations, introduced for tractability and to facilitate tuning. 
    
    \smallskip
The selection of a decision rule depends on how the mapping is judged during tuning and evaluation. This is a three-step procedure, whose first two steps are its design, involving $(i)$ selecting a type of decision rule $\mySet{F}$ (e.g., linear model, decision tree, \ac{dnn}, etc.); and $(ii)$ tuning its parameters $\myVec{\theta}$ based on the decision rule objective. Then $(iii)$, the tuned system is evaluated in simulations based on the evaluation objective.  The evaluation is determined by the simulator, and is independent of the design steps. Fig.~\ref{fig:DesignProc1} illustrates the overall procedure.
The traditional  approaches to carry out the design procedure, referred to as {\em model-based} or {\em classic} methods, are based on modelling and knowledge; the data-centric approach uses \ac{ml}, with deep learning being a leading family of \ac{ml} techniques.

\section{Model-Based Methods}
\subsection{Decision Rule Objective}
   The classic model-based approach sets decision rules based on domain knowledge. Namely, knowledge of an underlying model which mathematically describes the setup 
   is used along with the loss measure $l(\cdot)$ to formulate an analytical surrogate decision rule objective $\mySet{L}:\mySet{F}\mapsto \mySet{R}^+$. Both the model imposed and the objective are typically simplified approximations of the evaluation simulator and objective, respectively, introduced for analytical tractability. The decision rule objective also often includes sensitivity and regularization terms, resulting in an inductive bias on $f$. 
   The decision rule objective is applied to select the decision rule from $\mySet{F}$,  which can  be a pre-defined type or the entire space of mappings from $\InputSpace$ to $\LabelSpace$. 
   Once a simplified  objective $\mySet{L}$ is set, one can often find the optimal decision rule in $\mySet{F}$ with respect to $\mySet{L}$.  
   
   For instance, for inference tasks, given knowledge of a  distribution  $\Distribution$ defined over $\InputSpace \times \LabelSpace$, one can formulate the risk  $ \mySet{L}(f) =\E_{(\Input,\Label)\sim \Distribution} \{l(f,\Input,\Label)\}$, and set $f$ to minimize the error rate risk among all mappings from $\InputSpace$ to $\LabelSpace$ as the {\em \ac{map}} rule, given by:
    \begin{align}
        f_{\rm MAP}(\Input) 
        &=  \mathop{\arg \max}\limits_{\Output \in \LabelSpace} \Pr(\Label = \Output | \Input).
        \label{eqn:Map}
    \end{align}

    The formulation of  $ \mySet{L}(f)$ is dictated by the model imposed on the underlying relationship between $\Input$ and the desired  $\Label$. This objective typically contains parameters of the model, which we denote by $\ObjParam$, and henceforth write $\mySet{L}_{\ObjParam}(f)$. 
    
    \begin{example}
    \label{exm:InverseModel1}
    A common approach to treat the super-resolution problem in Example~\ref{exm:Inverse} is to assume the compression obeys a linear Gaussian model, i.e., 
    	\begin{equation}
    	\label{eqn:linearGuassian}
	    \myVec{x} = \myMat{H}\Label + \myVec{w}, \qquad \myVec{w} \sim \mathcal{N}(\myVec{0}, \sigma^2 \myMat{I}).
	\end{equation}
	The matrix $\myMat{H}$ in \eqref{eqn:linearGuassian} may represent the point-spread function of the system, a reduced measurement resolution, etc.
	The \ac{map} rule in \eqref{eqn:Map} becomes 
	\begin{equation}
	\label{eqn:Recovery1}
	     f_{\rm MAP}(\Input) = \mathop{\arg\min}\limits_{\myVec{s}} \frac{1}{2}\|\myVec{x}-\myMat{H}\myVec{s}\|^2_2 +\sigma^2\phi(\myVec{s}),
	\end{equation}
	where $\phi(\myVec{s}) := -\log \Pr(\Label = \myVec{s})$. The resulting decision rule objective requires imposing a prior on $\mySet{S}$ encapsulated in  $\phi(\myVec{s})$. A popular selection is to impose sparsity in some known domain $\myMat{\Psi}$ (e.g., wavelet), such that $\Output = \myMat{\Psi}\myVec{r}$, where $\myVec{r}$ is sparse. This boils down to an objective  defined on $\myVec{r}$,  given by
	\begin{equation}
	    \label{eqn:InvesObj1}
	    \mySet{L}_{ \ObjParam}(\myVec{r}) =  
	    \frac{1}{2}\|\myVec{x}-\myMat{H}\myMat{\Psi}\myVec{r}\|^2_2 +\rho\|\myVec{r}\|_0,
	\end{equation}
	where the parameter $\rho$ encapsulates  $\sigma^2$ and the expected sparsity level. The parameters of the   objective  in \eqref{eqn:InvesObj1} are 
	\begin{equation}
	\label{eqn:vPsi}
	    \ObjParam = \{\myMat{H},\myMat{\Psi},\rho\}.
	\end{equation}
    \end{example}
    
    The above example shows how one can leverage domain knowledge to formulate an objective, which is dictated by the parameter vector   $\ObjParam$. It also demonstrates  two key properties of model-based approaches: $(i)$ that surrogate models can be quite unfaithful to the true data, since, e.g., the Gaussianity of $\myVec{w}$ implies that $\myVec{x}$ in \eqref{eqn:linearGuassian} can take negative values, which is not the case for image data; and $(ii)$ that simplified models allow translating the task into a relatively simple closed-form objective, as in \eqref{eqn:InvesObj1}.  Similar approaches can be used to tackle the stochastic control setting of Example~\ref{exm:Control}.
    
    \begin{example}
    \label{exm:ControlModel1}
    Traditional \ac{lqg} control considers dynamics that take the form of a linear Gaussian state-space model, where
    \begin{subequations}
    \label{eqn:ssmodel}
    \begin{align}
        \myVec{z}_{t+1} &= \myMat{A}\myVec{z}_{t} + \myMat{B}\Output_t + \myVec{v}_t, \\
        \myVec{x}_t &=  \myMat{C}\myVec{z}_{t} +\myVec{w}_t.
        \label{eqn:ssmodelObs}
    \end{align}
    \end{subequations}
    Here, the noise sequences $\myVec{v}_t, \myVec{w}_t$ are zero-mean Gaussian signals, i.i.d. in time, with covariance matrices $\myMat{V}, \myMat{W}$, respectively. The objective at each time instance $t$ is given by
    \begin{equation}
	    \label{eqn:LQGObj1}
        \mySet{L}_{ \ObjParam}(f) = \E\{\myVec{z}_{t}^T \myMat{Q} \myVec{z}_{t} + \Output_{t}^T \myMat{R} \Output_{t}  \}, \quad \Output_t = f(\{\myVec{x}_{\tau}\}_{\tau \leq t}).
    \end{equation}
    The parameters of the objective function \eqref{eqn:LQGObj1} are thus
	\begin{equation}
	    \label{eqn:LQGObj2}
	    \ObjParam = (\myMat{A},\myMat{B},\myMat{C}, \myMat{Q}, \myMat{R}, \myMat{V}, \myMat{W}).
	\end{equation}
    \end{example}
    
      \begin{example}
    \label{exm:MPC}
    Model predictive control replaces the expectation based objective in \eqref{eqn:LQGObj1} with a deterministic optimization problem based on forecasting over some finite horizon $H$. Here, for the linear Gaussian state-space model of \eqref{eqn:ssmodel} with a quadratic loss, the objective at each time period $t$ is given by
    \begin{equation}
	    \label{eqn:MPCobj1}
        \mySet{L}_{ \ObjParam}(f) = \sum_{\tau=0}^{H-1}\left(\hat{\myVec{z}}_{t+\tau}^T \myMat{Q}_{\tau} \hat{\myVec{z}}_{t+\tau} + \Output_{t+\tau}^T \myMat{R}_{\tau} \Output_{t+\tau}  \right),
    \end{equation}
    where $\Output_t,\ldots,\Output_{t+H-1} = f(\{\myVec{x}_{\tau}\}_{\tau \leq t})$ and $\{\hat{\myVec{z}}_{t+\tau}\}$ are computed via \eqref{eqn:ssmodel} with $\{{\myVec{v}}_{t+\tau}\}$ and $\{{\myVec{w}}_{t+\tau}\}$ replaced with some predicted values. 
    The parameters $\ObjParam$ of the objective function \eqref{eqn:MPCobj1} thus include these predictive mapping, as well as the matrices $\myMat{A},\myMat{B},\myMat{C}, \{\myMat{Q}_\tau, \myMat{R}_\tau\}$.
	  \end{example}
    
    The formulation of the decision rule objectives in Examples~\ref{exm:InverseModel1}-\ref{exm:MPC} relies on full domain knowledge, e.g., one has to know the prior $\phi(\cdot)$ or the covariances $\myMat{V},\myMat{W}$ in order to express the objectives in \eqref{eqn:InvesObj1} and \eqref{eqn:LQGObj1}, respectively. 

\subsection{Decision Rule Type}
 Model-based methods determine the decision rule objective based on domain knowledge, obtained from measurements and from understanding of the underlying physics. 
 Once the objective is fixed, evaluating the decision rule boils down to solving an optimization problem, typically resulting in {\em highly-specific} types of decision mappings whose structure follows from the optimization formulation. In particular, a  decision rule is typically  obtained as either an {\em explicit solution} of the  problem, or in the form of an {\em iterative solver}.
    
    {\bf Explicit solvers} arise when the decision rule objective takes a relatively simplified form, such that one can characterize the optimal mapping. In such cases, the optimization-based decision rule of type~\ref{itm:opt} can specialize into an affine rule~\ref{itm:affine}.
    \begin{example}
    \label{exm:LQGPoliciy}
    The mapping which minimizes the  \ac{lqg}  loss in Example~\ref{exm:ControlModel1} is known to be obtained by first predicting the latent state $\myVec{z}_t$ using a Kalman filter, i.e., 
    \begin{equation}
    \hat{\myVec{z}}_t = \myMat{A}\hat{\myVec{z}}_{t-1} + \myMat{L}_t\left(\myVec{x}_t - \myMat{C}(\myMat{A}\hat{\myVec{z}}_{t-1}+  \myMat{B}\Output_{t-1}\right),
    \label{eqn:Kalman}
    \end{equation}
    where $\myMat{L}_t$ is the Kalman gain matrix. The action is taken to be
    \begin{equation}
        \Output_t = - \myMat{K}_t\hat{\myVec{z}}_t,
    \end{equation}
    with $\myMat{K}_t$ being the feedback gain matrix. Both $\myMat{L}_t$ and $\myMat{K}_t$  are determinstically determined by $\ObjParam$ in \eqref{eqn:LQGObj2}, and are updated based on internally tracked statistical moments that are recursively updated. 
    \end{example}
    
    Example~\ref{exm:LQGPoliciy} demonstrates how the modelling of a complex task using a simplified linear Gaussian model, combined with the usage of a simple surrogate quadratic objective, results in an explicit solution, which here takes a linear form. While this surrogate model and the  objective are likely to differ from the operation of the system, one can tune the objective parameters $\ObjParam$ (encapsulated in \eqref{eqn:Kalman} and in $\myMat{K}_t$) via simulations, thus modifying the decision rule to match the expected operation.
        
    {\bf Iterative solvers} follow  mathematical steps which gradually lead to the decision that achieves the decision rule objective, yielding a mapping as in type~\ref{itm:iter}.  A large body of optimization techniques are iterative, with common schemes based on first-order methods (i.e., gradient iterative steps) \cite[Ch. 9]{boyd2004convex}.  
    Iterative optimizers typically give rise to additional parameters which affect the speed and convergence rate of the algorithm, but not the actual objective being minimized. We refer to these parameters of the solver as {\em hyperparameters}, and denote them by $\HypParam$. As opposed to  the objective parameters $\ObjParam$ (as in, e.g., \eqref{eqn:vPsi}), they often have no effect on the solution when the algorithm is allowed to run to convergence, and so are of secondary importance.   But when the iterative algorithm is stopped after a predefined number of iterations, they affect the decisions, and therefore also the decision rule objective. Due to the surrogate nature of the  objective, such stopping does not necessarily degrade the evaluation performance.
    \begin{example}
    \label{exm:ADMM}
    The super-resolution objective in \eqref{eqn:Recovery1} can be tackled 
		using the \ac{admm} \cite{boyd2011distributed}. This method summarized as Algorithm~\ref{alg:Algoadmm}, where we merge $\sigma^2$ in \eqref{eqn:Recovery1} into the prior $\phi(\cdot)$ for brevity, and the proximal mapping is defined as 
		\begin{equation}
		 \label{eqn:proxi}
		 {\rm prox}_g(\myVec{v}):= \arg\min_{\myVec{z}} \left( g(\myVec{z})+\frac{1}{2}\|\myVec{z}-\myVec{v}\|^2_2 \right).
		\end{equation}
    \end{example}
    \vspace{-0.4cm}
    
  \begin{algorithm}  
    \caption{\ac{admm}}
    \label{alg:Algoadmm}
    \begin{algorithmic}
    		\State Fix step size $\mu$, and  $\lambda>0$. 
    		\State Initialize $k=0$, $\myVec{u}_0$,  $\myVec{v}_0$ randomly.
    		\Repeat
        		\State    Update $
        	    \myVec{s}_{k+1} = (\csMatrix^T\csMatrix  + 2\lambda \myMat{I})^{-1}(\csMatrix^T\myVec{x} + 2\lambda(\myVec{v}_k - \myVec{u}_k))$.
        	    \State \label{stp:prox}
        	    Update  $
        	    \myVec{v}_{k+1}  
        	    =  {\rm prox}_{\frac{1}{2\lambda}\phi}(\myVec{s}_{k+1} + \myVec{u}_k)$ (see \eqref{eqn:proxi}).
        	    \State    	    Update  $
        	    \myVec{u}_{k+1}  
        	    = \myVec{u}_{k} +\mu \left(\myVec{s}_{k+1} -\myVec{v}_{k+1}\right)$. 
        	   \State Set $k=k+1$.
    	    \Until{convergence}
    		\State{Set estimate $\Output = \myVec{s}_{k}$.}
    \end{algorithmic} 
    \end{algorithm} 
    
    \ac{admm} converges to a solution of  \eqref{eqn:Recovery1} when $\phi(\cdot)$ is convex, for any positive value of $\mu$.   When $\phi(\cdot)$ is not convex, there are no convergence guarantees, but it has been observed in practice that good results are obtained, when $\mu$ is chosen appropriately.
     Since convergence of iterative optimizers to an optimal decision can be generally guaranteed for convex objectives, one often has to relax and modify the objective. 
\begin{example}
    \label{exm:ISTA}
    The non-convex super-resolution surrogate objective with sparse prior in \eqref{eqn:InvesObj1}  can be relaxed into 
    	\begin{equation}
	\label{eqn:LASSO} 
	    \mySet{L}_{ \ObjParam}(\myVec{s}) = 
	    \frac{1}{2}\|\myVec{x}-\myMat{H}\myVec{s}\|^2_2 +\rho\|\myVec{r}\|_1, 
	\end{equation}
	where we set $\myMat{\Psi}$ to the identity matrix for simplicity. This successive relaxation of an already surrogate objective  yields a convex cost  in \eqref{eqn:LASSO}. It can be solved, e.g., using proximal gradient descent with step size $\mu$, which specializes here into the {\em \ac{ista}} \cite[Ch. 7]{parikh2014proximal}. Letting  $\mySet{T}_{\beta}(\cdot) \triangleq {\rm sign}(x)\max(0,|x|-\beta)$ be the element-wise soft-thresholding operation, the  update equation is 
    \begin{equation}
    \label{eqn:ISTA}
         \myVec{s}_{k+1} = \mySet{T}_{\mu\rho}\left( \myVec{s}_{k} + \mu \myMat{H}^T(\myVec{x}-\myMat{H}\myVec{s}_k) \right).
    \end{equation} 
\end{example}
    
    In Example \ref{exm:ADMM}, illustrated in Fig.~\ref{fig:ADMMComp}(a), the iterative solver introduces two hyperparameters, i.e., $\HypParam = [\lambda, \mu]$, which are used in the iterative minimization of \eqref{eqn:Recovery1}. In Example~\ref{exm:ISTA}, there is only one hyperparameter  $\HypParam = \mu$.   The hyperparameters $\HypParam$ are often set by manual hand tuning based on simulations.

    \subsection{Summary}
    Model-based methods rely on   decision rules of type~\ref{itm:opt}, where an  analytically tractable optimization problem is formulated based domain knowledge. 
 The optimization problems solved are typically {\em surrogates} for the real application problem.   The decision rule  objectives and constraints are often inspired by physical characteristics, understanding of the system operation, and existing models (of noise, disturbances, and other  quantities). Yet, in practice, objectives are likely to differ from the system task, due to multiple reasons, including:
    \begin{itemize}
        \item Simplifying approximations, e.g., modelling super-resolution as a linear Gaussian setup in Example~\ref{exm:InverseModel1}.
        \item Estimation inaccuracies, e.g., substituting estimated covariances $\myMat{W},\myMat{V}$ to compute the \ac{lqg} objective in Example~\ref{exm:ControlModel1} or an estimated  $\myMat{H}$ in the \ac{map} objective in Example~\ref{exm:InverseModel1}.
        \item Introducing regularization terms in the objective, e.g., $\rho \|r\|_0$ in  Example~\ref{exm:InverseModel1}.
        \item Relaxations or approximations of the objective and constraints to render the  optimization problem solvable.
       \item Scaling of some of the quantities involved.
    \end{itemize}
    
    Model-based techniques are particularly suitable for the resulting optimization problem. Once a solvable (e.g., convex) formulation is determined, these methods are guaranteed to obtain its solution. Furthermore, their operation is interpretable, and tends to be highly flexible, as one can substitute different values of the objective parameters $\ObjParam$.
    
     In practice, accurate knowledge of the statistical model relating the context and the desired decision is often unavailable. Consequently, model-based techniques may require imposing  assumptions on the underlying statistics, which  in some cases  reflect the actual behavior, but can also be a crude approximation. In the presence of inaccurate model knowledge,   either as a result of estimation errors or due to enforcing a model which does not fully capture the environment, the performance of model-based techniques tends to degrade during evaluation. This limits the applicability of model-based schemes in scenarios where one cannot represent the task via a decision rule objective in a closed-form (and preferably simplified) expression, or alternatively, when the underlying model is costly to estimate accurately, or too complex to express analytically. Additional challenges stem from the fact that decision making can be slow, particularly  using iterative solvers. Finally, setting the hyperparameters $\HypParam$  is often elusive, and may involve heuristics and cumbersome hand-crafted tunning. 

\section{Deep Learning}
\subsection{Decision Rule Objective} 
   While in many applications coming up with accurate and tractable statistical modelling is difficult, we are often given access to data describing the setup. 
   \ac{ml} systems learn their mapping from data. In a supervised setting, data is comprised of a set  of $\Ntraining$ pairs of inputs and labels, denoted $\Data = \{\Input_i, \Label_i\}_{i=1}^{\Ntraining}$. In reinforcement learning, data is obtained from a simulator, which on each decision produces a subsequent context. This data is referred to as the {\em training set}, and there is typically an additional data set used for evaluation and validation.
    Since no mathematical model relating the input and the desired decision is imposed, the decision rule objective is often the {\em empirical risk}. Focusing on a supervised setting, this objective is given by 
   \begin{equation}
   \label{eqn:EmpRisk}
       \mySet{L}_{\Data}(f) \triangleq \frac{1}{\Ntraining}\sum_{i=1}^{\Ntraining} l(f,\Input_i, \Label_i).
   \end{equation}
   
   Decision rule objectives that are based on data and do not rely on modeling are often a more faithful representation of the evaluation objective compared with model-based approaches. However, they are still surrogates. This follows not only from the difference between the training and  validation data, but also from the frequent inclusions of regularizing terms  and  mechanisms such as dropout and batch normalization, whose operation differs between training and evaluation. 
   
   While we focus our description on supervised settings, \ac{ml} systems can also learn in an unsupervised manner. In such cases, the data set $\Data$ is comprised only of a set of examples $\{\Input_i\}_{i=1}^{\Ntraining}$, and the loss measure $l$ is defined over $\mathcal{F}\times\InputSpace$, instead of over $\mathcal{F}\times \InputSpace \times \LabelSpace$ as  in \eqref{eqn:lossfunction}.  
	Unsupervised \ac{ml} is often used to discover  patterns in the  data, with tasks including clustering, anomaly detection, generative modeling, and compression.


\subsection{Decision Rule Type} 
    In contrast to the model-based case, where decision rules can sometimes be derived by directly solving the optimization problem  without initially imposing structure on the system, setting a decision rule based on data necessitates restricting the domain of feasible mappings. This stems from the fact that  one can usually form a decision rule which minimizes the empirical loss of \eqref{eqn:EmpRisk} by memorizing the data, i.e., overfit  \cite[Ch. 2]{shalev2014understanding}.     A leading strategy in \ac{ml}, upon which deep learning is based, is to assume a highly-expressive generic parametric model on the decision mapping, while incorporating optimization mechanisms and regularizing the empirical risk to avoid overfitting. 
    In deep learning, $f$ is a \ac{dnn}, i.e., type~\ref{itm:dnn}, with the parameters $\myVec{\theta}$ being the network parameters, e.g., the
    weights and biases of each layer. By the universal approximation theorem, \acp{dnn} can  approach any Borel measurable mapping \cite[Ch. 6]{goodfellow2016deep}.

	\begin{figure*}
	    \centering
	    \includegraphics[width=\linewidth]{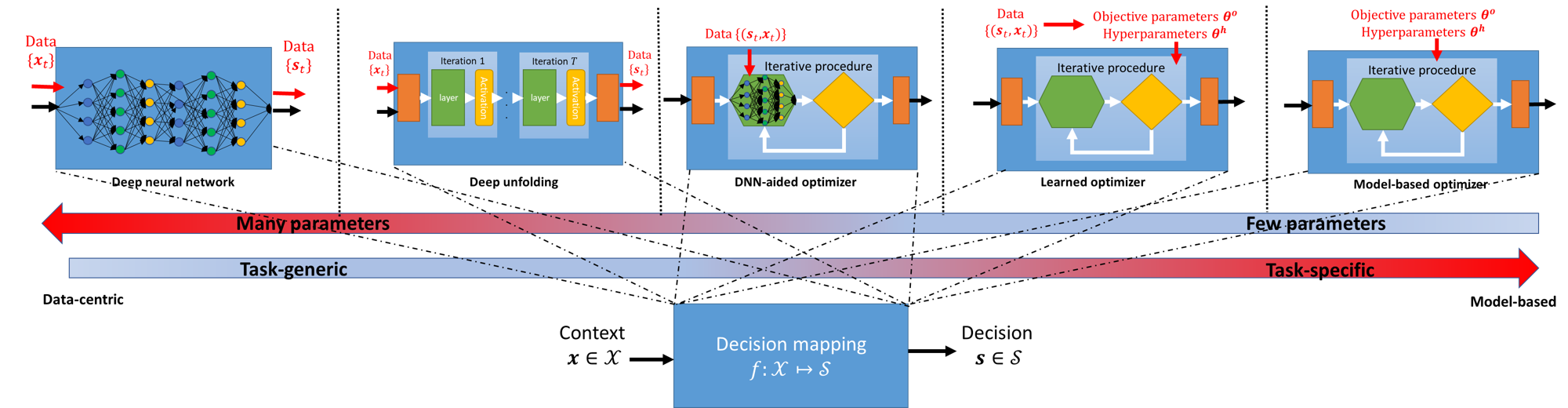}
	    \caption{Continuous spectrum of specificity and parameterization with model-based methods and deep learning constituting the extreme edges of the spectrum.}
	    \label{fig:Spectrum1}
	    \vspace{-0.2cm}
	\end{figure*}	

While model-based algorithms are specifically tailored to a given scenario, deep learning is model-agnostic. The unique characteristics of the  scenario are encapsulated in 
 the learned weights. The decision rule family $\mySet{F}$, i.e., the possible \ac{dnn} mappings, is generic and can be applied in a broad range of different problems.  
While standard \ac{dnn} structures are model-agnostic and are commonly treated as black boxes, one can still incorporate some level of domain knowledge in the selection of the network architecture. For instance, when the input is known to exhibit temporal correlation, architectures based on \acp{rnn} \cite[Ch. 10]{goodfellow2016deep} or attention mechanisms \cite{vaswani2017attention} are often preferred. Alternatively, in the presence of spatially local patterns, one may   utilize convolutional layers \cite{lecun1995convolutional}. An additional method to incorporate domain knowledge into a black box \ac{dnn} is by pre-processing of the input via, e.g., hand-crafted feature extraction. 

\begin{example}
    The super-resolution task (Example~\ref{exm:Inverse}) can be carried out by training a deep convolutional autoencoder \cite{mao2016image}.
\end{example}

\begin{example}
    Stochastic control (Example~\ref{exm:Control}) can be carried by a \ac{dnn} controller using deep reinforcement learning \cite{lillicrap2015continuous}.
\end{example}

    The fact that \acp{dnn} are comprised of a large number of  parameters, and that massive data sets are often used for their training, makes it unlikely to recover $\myVec{\theta}$ that minimizes the empirical risk with affordable computational effort. Instead, the tuning of $\myVec{\theta}$ is typically carried out using first-order gradient-based algorithms, where gradients  estimated from a small number of randomly chosen samples, e.g., by mini-batch \ac{sgd} iterations of the form
			\begin{equation}
			\label{eqn:SGD1}
			\myVec{\theta}_{j+1}= \myVec{\theta}_{j} - \eta_j \nabla_{\myVec{\theta}} \mySet{L}_{\mySet{D}_j}(f(\cdot;{\myVec{\theta}_j})),
			\end{equation} 
    where $\mySet{D}_j$ is a mini-batch sampled  from $\mySet{D}$, and $\eta_j$ is the learning rate. The gradients in \eqref{eqn:SGD1} are computed using backpropagation \cite{rumelhart1985learning}. Mini-batch \ac{sgd}  is the basis for \ac{dnn} training, with common variants using momentum and adaptive learning rates. Such training methods operate in an automated manner, enabling tuning of \acp{dnn} from massive data sets.

\subsection{Summary}
\acp{dnn} operate in a model-agnostic manner, and can be tuned to implement an immense family of mappings, making them widely adopted in areas where principled mathematical models are scarce, such as computer vision and natural language processing.
Despite their success, existing deep learning approaches are subject to several challenges, which limit their applicability in some application domains. The computational burden of training and utilizing highly parametrized \acp{dnn}, as well as the fact that massive data sets are often required for their training,  constitute major drawbacks in various signal-processing, communications and in control applications. 
This limitation is particularly relevant when operating on hardware-constrained devices, e.g., mobile systems, unmanned aerial vehicles, and sensors. Such systems are typically limited in their ability to utilize highly parametrized \acp{dnn},  and they should be flexible to adapt to variations in the environment. Furthermore, the fact that the decision mapping is learned solely from data often gives rise to generalization issues  on unseen data.
Finally, due to the complex and generic structure of \acp{dnn}, it is often extremely challenging to understand how they obtain their predictions, track the rationale leading to their decisions, and characterize confidence intervals. Consequently, deep learning does not offer the interpretability, flexibility, versatility, and reliability of model-based methods. This is a major limitation for tasks involving  critical and even life-saving decision making, such as the control of vehicular and aerospace systems. 

\section{Hybrid Model-Based Deep Learning Optimizers}
Model-based methods and deep learning are often viewed as fundamentally different approaches for setting decision boxes. Nonetheless, both strategies typically use parametric mappings, i.e., the weights  $\myVec{\theta}$ of \acp{dnn} and the parameters $(\ObjParam,\HypParam)$ of model-based optimizers, whose setting is determined based on data and on  principled mathematical models. The core difference thus lies in the specificity and the parameterization of the decision rule type:  Model-based methods are {\em knowledge-centric}, using decision rules that are  task-specific, and usually involve a limited number of parameters that can often be set manually. Deep learning is {\em data-centric}, and thus uses highly-parametrized model-agnostic task-generic mappings. 

The identification of model-based methods and deep learning as two ends of a spectrum of specificity and parameterization indicates the presence of a continuum, as illustrated in Fig.~\ref{fig:Spectrum1}. In fact, many techniques lie in the middle ground, designing decision rules with different levels of specificity and parameterization by combining some balance of deep learning  with model-based optimization \cite{chen2021learning}. In this section we review three systematic frameworks for designing decision mappings that are both knowledge- and data-centric as a form of hybrid model-based deep learning: The first strategy, coined {\em learned optimizers} \cite{agrawal2021learning, agrawal2020learning}, uses deep learning  automated tuning machinery to tune parameters of model-based optimization conventionally tuned by hand. The second family of techniques, referred to as {\em deep unfolding} \cite{monga2021algorithm}, converts iterative optimizers into \acp{dnn}. The third type of model-based deep learning schemes, which we call {\em \ac{dnn}-aided optimizers} \cite{shlezinger2020model}, augments model-based optimization with dedicated \acp{dnn}.

\subsection{Learned Optimization}
Learned optimizers use conventional model-based methods for decision making, while tuning the parameters and hyperparameters of classic solvers via automated deep learning training \cite{agrawal2021learning, agrawal2020learning}. This form of model-based deep learning leverages data to optimize the optimizer. While learned optimization bypasses the traditional daunting effort of manually fitting the decision rule parameters, it involves the introduction of new hyperparameters of the training procedure that must to be configured (typically by hand). 

Learned optimization effectively converts an optimizer into an \ac{ml} model. Since automated tuning of \ac{ml} models typically uses gradient-based methods as in \eqref{eqn:SGD1}, a key requirement is for the optimizer to be differentiable, namely, that one can compute the gradient of its decision with respect to its parameters. Fortunately, convex optimization solvers are typically differentiable (under some regularity conditions) \cite{agrawal2019differentiable}. Alternatively, for non-convex optimization, one can differentiate numerically \cite{maclaurin2015gradient} or, in some cases,  implicitly~\cite{lorraine2020optimizing}.

{\bf Examples:}
Learned optimization focuses on optimizing parameters conventionally tuned manually; these are parameters whose value does not follow from prior knowledge of the problem being solved, and thus their modification affects only the solver, and not the problem being solved. For model-based optimizers based on explicit solutions, the parameters available are only those of the objective $\ObjParam$. Nonetheless, some of these parameters stem from the fact the objective is inherently a surrogate for the actual problem being solved, and thus require tuning, as shown in the following example.

\begin{example}
    \label{exm:BPKalman}
    Consider a dynamic system characterized by a state-space model as in \eqref{eqn:ssmodel}. In such  settings, the linear mappings $\myMat{A},\myMat{B}, \myMat{C}$ often arise from understanding the physics of the problem, while the objective parameters $\myMat{Q}$ and $\myMat{R}$ stem from the system requirements. Nonetheless, in practice, one typically does not have a concrete stochastic model for the noise signals, which are often introduced as a way to capture stochasticity, and thus $\myMat{V}$ and $\myMat{W}$ are often tuned by hand. 
    
    Given a data set of $\Ntraining$ trajectories of $T$ observations  with the corresponding states and actions $\mySet{D}=\{\{\myVec{x}_{t,i}, \Output_{t,i}, \myVec{z}_{t,i}\}_{t=1}^T\}_{i=1}^{\Ntraining}$, one set the trainable parameters to be $\myVec{\theta}=[\myMat{V},\myMat{W}]$, and optimize them via \eqref{eqn:SGD1} with $\mySet{L}_{\mySet{D}}$ being the empirical $\ell_2$ distance between the Kalman filter prediction \eqref{eqn:Kalman} and the true state \cite{barratt2020fitting}. The gradients of $\mySet{L}_{\mySet{D}}$ are computed using \ac{bptt} \cite{sutskever2013training},  building upon the diffrentiability of the Kalman gain $\myMat{L}_t$ with respect to both $\myMat{V}$ and $\myMat{W}$ \cite{xu2021ekfnet}. 
\end{example}

When the optimizer being learned is an iterative solver, one can use data to tune the hyperparameters $\HypParam$, whose value does not affect the optimization objective. 
This can be specialized for the \ac{admm} optimizer of Example~\ref{exm:ADMM}, as shown next.

\begin{figure*}
    \centering
    \includegraphics[width=\linewidth]{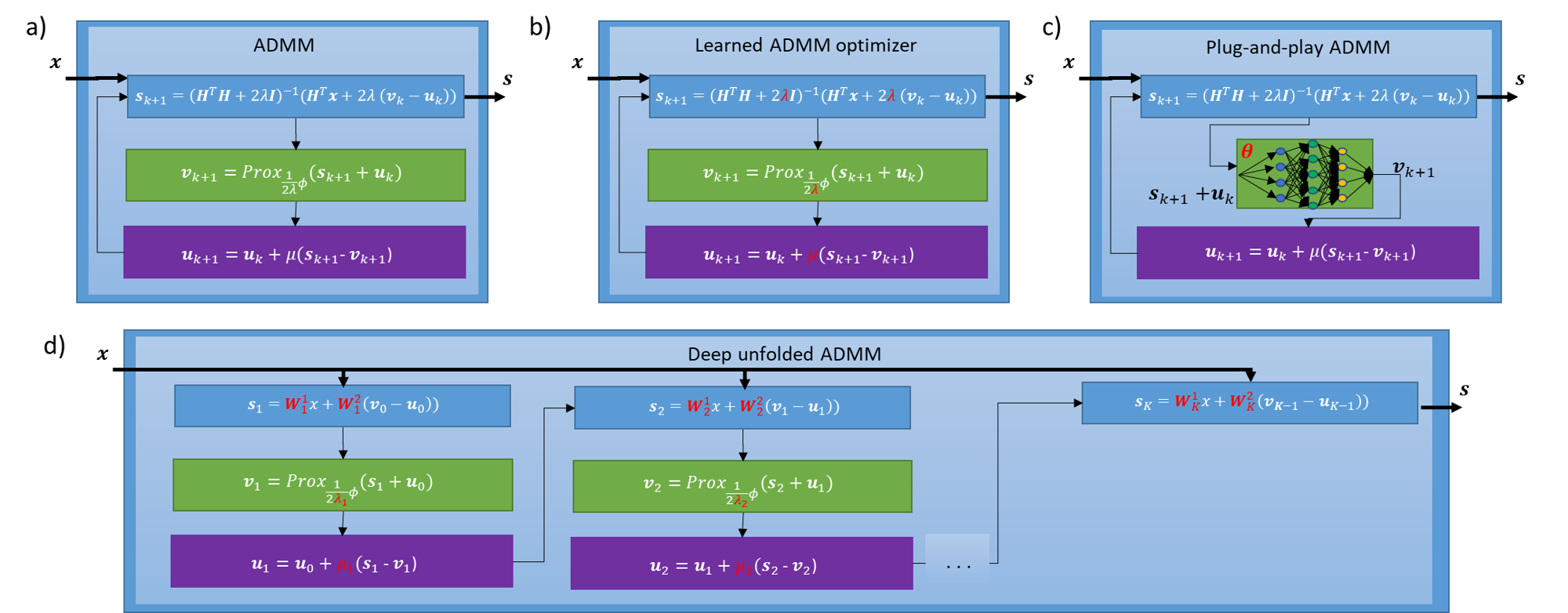}
    \caption{An illustration of the model-based deep learning strategies arising from the \ac{admm} optimizer (Algorithm~\ref{alg:Algoadmm}), where variables in \textcolor{red}{red fonts} represent trainable parameters: $a)$ the model-based optimizer (Example~\ref{exm:ADMM}); $b)$ a learned \ac{admm} optimizer (Example~\ref{exm:BPADMM}); $c)$ plug-and-play \ac{admm} (Example~\ref{exm:PnPADMM}); and $d)$ deep unfolded \ac{admm} (Example~\ref{exm:DeepUnfADMM}).}
    \label{fig:ADMMComp}
    \vspace{-0.2cm}
\end{figure*}

\begin{example}
    \label{exm:BPADMM}
    Consider the \ac{admm} solver (Algorithm~\ref{alg:Algoadmm}). Given a data set $\mySet{D}$ of $\Ntraining$ labeled samples, the hyperparameter vector $\HypParam = [\lambda,\mu]$ can be optimized by treating it as trainable parameters, as visualized in Fig.~\ref{fig:ADMMComp}(b). Letting $f(\cdot; \HypParam)$ be the \ac{admm} mapping with hyperparameters $\HypParam$, this is given by
    \begin{equation}
        \myVec{\theta}^* = \mathop{\arg \min}_{\HypParam=[\lambda, \mu] \in \mySet{R}^+} \frac{1}{\Ntraining}\sum_{i=1}^{\Ntraining}\|f(\myVec{x}_i;\HypParam) - \Label_i\|^2_2.
        \label{eqn:BPAdmmObj}
    \end{equation}
    The problem in \eqref{eqn:BPAdmmObj} is as \ac{dnn} training, e.g., \eqref{eqn:SGD1}, where to compute each gradient of the objective with respect to $\HypParam$, Algorithm~\ref{alg:Algoadmm} must first run until it reaches convergence, after which the gradients are computed via \ac{bptt}.  
\end{example}
 
{\bf Summary.} Learned optimizers are \ac{ml} decision rules which completely preserve the operation of conventional model-based methods. As such, they share the core gains of principled optimization. These include its suitability for the problem at hand; the interpretability that follows from the ability to relate each feature involved to an operation meaning; and flexibility, as one can control the objective for which the decision is configured using its non-learned parameters in $\ObjParam$.  

Compared with model-based optimization, learned optimizers facilitate the design procedure, avoiding the need to tune parameters by hand. Furthermore, the fact that tuning is carried out by observing the decision output and evaluating it based on data allows to improve performance when the surrogate objective differs from the (possibly analytically intractable) evaluation objective. Finally, when the decision box is an iterative solver, learned optimization can reduce the convergence speed compared with manually tuned $\HypParam$.

\subsection{Deep Unfolding}
A relatively common  methodology for combining model-based methods and deep learning is that of deep unfolding, also referred to as {\em deep unrolling} \cite{monga2021algorithm}. Originally proposed by Greger and LeCun for sparse recovery \cite{gregor2010learning}, deep unfolding converts iterative optimizers into trainable \acp{dnn}. As the name suggests, the method unfolds an iterative algorithm into a sequential procedure with a fixed number of iterations. Then, each iteration is treated as a layer, with its trainable parameters $\myVec{\theta}$ being either only the hyperparameters $\HypParam$, or also the decision rule objective parameters $\ObjParam$.

Unfolding an iterative optimizer into a \ac{dnn} facilitates tuning different parameters for each iteration, being converted into trainable parameters of different layers. This is achieved by training  end-to-end, i.e., by evaluating the system output based on data. Letting $K$ be the number of unfolded iterations, deep unfolding can learn iteration-dependent hyperparameters $\{\HypParam_k\}_{k=1}^K$ and even objective parameters $\{\ObjParam_k\}_{k=1}^K$. This increases the parameterization and abstractness compared with learned optimization of iterative solvers, which typically reuses the learned hyperparameters and runs until convergence (as in model-based optimizers). Nonetheless, for every setting of $\{\ObjParam_k\}_{k=1}^K$ and $\{\HypParam_k\}_{k=1}^K$, a deep unfolded system effectively carries out its decision using $K$ iterations of some principled iterative solver known to be suitable for the problem.

{\bf Examples:} 
Deep unfolded networks can be designed to improve upon model-based optimization in convergence speed and model abstractness. The former is achieved since the resulting system operates with a fixed number of iterations, which can be  much smaller compared with that usually required to converge. This is combined with the natural ability of deep unfolding to learn iteration-dependent hyperparameters to enable accurate decisions to be achieved within this predefined number of iterations, as exemplified next: 
\begin{example}
\label{exm:UnfADMM}
Let us consider again the \ac{admm} optimizer of Algorithm~\ref{alg:Algoadmm}. A deep unfolded \ac{admm} is obtained by setting the decision to be $\Output = \myVec{s}_K$ for some fixed $K$, and allowing each iteration to use hyperparameters $[\lambda_k, \mu_k]$, that are stacked into the trainable parameters vector $\myVec{\theta}$. Similarly to \eqref{eqn:BPAdmmObj}, these hyperparameters are learned from data via
 \begin{equation}
        \myVec{\theta}^* = \mathop{\arg \min}_{\myVec{\theta}=\{\lambda_k, \mu_k\}_{k=1}^K} \frac{1}{\Ntraining}\sum_{i=1}^{\Ntraining}\|f(\myVec{x}_i; \myVec{\theta}) - \Label_i\|^2_2.
        \label{eqn:UnfAdmmObj}
    \end{equation}
\end{example}

Example~\ref{exm:UnfADMM}  implements $K$ \ac{admm} iterations, as only the hyperparameters are learned. One can also transform iterative solvers into more abstract \acp{dnn} by also tuning the objective of each iteration. Continuing along the line of Examples~\ref{exm:BPADMM}-\ref{exm:UnfADMM}, we show how this is can be achieved following \cite{johnston2021admm}:
\begin{example}
\label{exm:DeepUnfADMM}
An alternative approach to unfold the \ac{admm} optimizer into a \ac{dnn} is by repeating the procedure in Example~\ref{exm:UnfADMM}, where algorithm is unfolded into $K$ iterations with each assigned  hypereparameters $[\lambda_k, \mu_k]$. Furthermore, the first update step of Algorithm~\ref{alg:Algoadmm} is now replaced with
\begin{equation}
\label{eqn:ADMMNet}
	    \myVec{s}_{k+1} = \myMat{W}_k^1\myVec{x} + \myMat{W}_k^2(\myVec{v}_k - \myVec{u}_k).
\end{equation}
For $\myMat{W}_k^1 =  (\csMatrix^T\csMatrix  + 2\lambda \myMat{I})^{-1}\csMatrix^T$ and $\myMat{W}_k^2 = 2\lambda (\csMatrix^T\csMatrix  + 2\lambda \myMat{I})^{-1}$, \eqref{eqn:ADMMNet} coincides with the corresponding step in Algorithm~\ref{alg:Algoadmm}. 
The trainable parameters $\myVec{\theta}=\{\myMat{W}_k^1, \myMat{W}_k^2,\lambda_k, \mu_k\}_{k=1}^K$ are learned from data by jointly minimizing
 \begin{equation}
         \myVec{\theta}^*=\mathop{\arg \min}_{\myVec{\theta}=\{\myMat{W}_k^1, \myMat{W}_k^2,{\lambda}_k,\mu_k\} } \frac{1}{\Ntraining}\sum_{i=1}^{\Ntraining}\|f(\myVec{x}_i; \myVec{\theta}) - \Label_i\|^2_2.
        \label{eqn:DeepUnfAdmmObj}
    \end{equation}
\end{example}

The resulting decision mapping of Example~\ref{exm:DeepUnfADMM} is illustrated in Fig.~\ref{fig:ADMMComp}(d).
Unlike Example~\ref{exm:UnfADMM} which only learns the hyperparameters, the unfolded \ac{admm} in Example~\ref{exm:DeepUnfADMM} jointly learns the hyperparameters and the objective parameters $\ObjParam$ per each iteration. This can be viewed as if each iteration  follows a different objective, such that the output after $K$ iterations most accurately matches the desired value. 
While each layer  in Example~\ref{exm:DeepUnfADMM}  has different parameters, one can enforce identical parameters across layers. 
The \ac{dnn} can realize a larger family of mappings compared with the original model-based optimizer, which serves as a principled initialization for the system, rather than its fixed structure as in Example~\ref{exm:UnfADMM}. 

A popular application of deep unfolding, which follows the rationale of Example~\ref{exm:DeepUnfADMM} with both $\ObjParam$ and $\HypParam$  learned end-to-end, is the unfolding of \ac{ista} into \ac{lista}~\cite{gregor2010learning}. 

\begin{example}
\label{exm:LISTA}
The \ac{ista} optimizer in Example~\ref{exm:ISTA} can be unfolded into the \ac{lista} \ac{dnn} architecture by fixing $K$ iterations and replacing the update step in \eqref{eqn:ISTA} with 
    \begin{equation}
    \label{eqn:LISTA}
         \myVec{s}_{k+1} = \mySet{T}_{\lambda_k}\left( \myMat{W}_k^1\myVec{x} + \mu_k \myMat{W}_k^2  \myVec{s}_{k} \right).
    \end{equation} 
For $\myMat{W}_k^1 = \mu\csMatrix^T$, $\myMat{W}_k^2 = \myMat{I} -\mu\csMatrix^T\csMatrix$, $\mu_k = 1$ and $\lambda_k=\mu \rho$, \eqref{eqn:LISTA} coincides with  model-based \ac{ista}.  
The trainable parameters $\myVec{\theta}=\big[\{\myMat{W}_k^1, \myMat{W}_k^2, \mu_k\}_{k=1}^K\big]$ are learned from data via end-to-end training as in \eqref{eqn:DeepUnfAdmmObj}.
\end{example}

{\bf Summary:} Deep unfolding designs dedicated \acp{dnn} whose architecture follows iterative optimization algorithms. Compared with conventional \acp{dnn} applied to similar tasks, deep unfolded networks are more task-specific and less parameterized, as the setting of their trainable parameters and their interconnection is based on a iterative solver suitable for such problems. As a result, deep unfolded networks tend to require less data for training compared with standard \acp{dnn}, and often achieve improved performance and generalization \cite{monga2021algorithm}. Furthermore, deep unfolded networks offer improved interpretability, as one can identify the meaning of some of its internal features, a task which is rarely achievable in conventional \acp{dnn}. In deep unfolded networks, the features exchanged between its layers represent the output of each iteration as in type~\ref{itm:iter}, and can thus be associated with an estimate of the decision which is gradually refined as  in iterative optimization. 

Compared with model-based optimization, converting an iterative solver (\ref{itm:iter}) into a \ac{dnn} with $K$ layers (\ref{itm:dnn}) typically results in faster inference. The fact that iteration-specific parameters are learned  end-to-end  allows deep unfolded networks to operate with much fewer layers compared with the number of iterations required by the model-based optimizer to achieve similar performance. Furthermore, the increased parameterization improves the abstractness of the decision rule, particularly when both the hyperparameters $\HypParam$ and the objective parameters $\ObjParam$ are jointly learned as in Examples~\ref{exm:DeepUnfADMM} and \ref{exm:LISTA}. Such unfolded networks  depart from the iterative algorithm from which they originates, allowing them to overcome mismatches and approximation errors associated with the need to specify a mathematically tractable surrogate objective for decision making. 
In particular, training an unfolded network designed with a mismatched model using data corresponding to the true underlying scenario typically yields improved performance compared to the model-based iterative algorithm with the same model-mismatch, as the unfolded network can learn to compensate for this mismatch~\cite{khobahi2021model}.

\subsection{DNN-Aided Optimization}
The third model-based deep learning strategy combines conventional \ac{dnn} architectures with model-based optimization to enable the latter to operate reliably in complex domains. The rationale here is to preserve the  objective and structure of a model-based decision mapping suitable for the problem at hand based on the available domain knowledge, while augmenting computations that rely on approximations and missing domain knowledge with model-agnostic \acp{dnn}. \ac{dnn}-aided optimizers thus aim at benefiting from the best of both worlds by accounting in principled manner for the available domain knowledge while using deep learning to cope with the elusive aspects of the problem description.

Unlike the aforementioned strategies of learned optimizers and deep unfolding, which are relatively systematic and can be viewed as recipe-style methodologies, \ac{dnn}-aided optimization accommodates a broad family of different techniques for augmenting model-based optimizers with \acp{dnn}. We next discuss some representative \ac{dnn}-aided optimization approaches.

{\bf Examples:} 
The straight-forward application of \ac{dnn}-aided optimization replaces an internal computation of a model-based solver with a dedicated \ac{dnn}, converting it into a trainable  model-based deep learning system.  An example of how this is done,   based on \cite{revach2021kalmannet}, is detailed next.
\begin{example}
\label{exm:KNet}
Consider again the setting of Example~\ref{exm:BPKalman}, where a Kalman filter is designed without knowing the distribution of the noise signals in \eqref{eqn:ssmodel}. Since the dependency on the noise statistics in the Kalman filter 
is encapsulated in the Kalman gain $\myMat{L}_t$, its computation can be replaced with a trainable \ac{dnn}, and thus \eqref{eqn:Kalman} is replaced with 
    \begin{equation}
    \hat{\myVec{z}}_t = \myMat{A}\hat{\myVec{z}}_{t-1} + h_{\myVec{\theta}}(\myVec{x}_t,\Output_{t-1} )\left(\myVec{x}_t - \myMat{C}(\myMat{A}\hat{\myVec{z}}_{t-1}+  \myMat{B}\Output_{t-1}\right)),
    \label{eqn:KalmanNEt}
    \end{equation}
where $h_{\myVec{\theta}}$ is a \ac{dnn} with parameters $\myVec{\theta}$. Particularly, since $\myMat{L}_t$ is updated recursively, its learned computation is carried out with an \ac{rnn}. By letting $f(\cdot;\myVec{\theta})$ be the latent state estimate computed using \eqref{eqn:KalmanNEt} with parameters $\myVec{\theta}$,  the overall system is trained end-to-end via
 \begin{equation}
         \myVec{\theta}^*=\mathop{\arg \min}_{\myVec{\theta} } \frac{1}{\Ntraining T}\sum_{i=1}^{\Ntraining}\sum_{t=1}^{T}\|
         f(\myVec{x}_{t,i},\Output_{t-1,i};\myVec{\theta}) - \myVec{z}_{t,i}\|^2_2.
        \label{eqn:KalmanNEtObj}
    \end{equation}
\end{example}

Example~\ref{exm:KNet} was shown in \cite{revach2021kalmannet} to overcome non-linearities and mismatches in the state-space model, outperforming the classical Kalman filter while retaining its data efficiency and interpretability. It is emphasized though that Example~\ref{exm:KNet} represents one approach to combine Kalman filtering with \ac{dnn}-aided optimization methodology. Additional techniques include the usage of an external \ac{dnn} operating in parallel with the filter and providing correction terms, as proposed in \cite{satorras2019combining}, and the application of the Kalman filter to learned features extracted by a \ac{dnn} as in \cite{coskun2017long}, exemplified next.
\begin{example}
\label{exm:KLearnedFetaures}
Consider a state-space model as in \eqref{eqn:ssmodel} where the observations $\Input_t$ are complex and non-linear, i.e., \eqref{eqn:ssmodelObs} does not hold. One can still apply a Kalman filter designed for a linear Gaussian setting by applying a \ac{dnn} $h_{\myVec{\theta}}(\cdot)$ to transform $\Input_t$ into features that follow the state-space model assumed by the model-based filter. Here,  \eqref{eqn:Kalman}  becomes 
    \begin{equation}
    \hat{\myVec{z}}_t = \myMat{A}\hat{\myVec{z}}_{t-1} + \myMat{L}_t\left(h_{\myVec{\theta}}(\myVec{x}_t) - \myMat{C}(\myMat{A}\hat{\myVec{z}}_{t-1}+  \myMat{B}\Output_{t-1}\right)),
    \label{eqn:KalmanNEt2}
    \end{equation}
    and the tuning is done via end-to-end training as in \eqref{eqn:KalmanNEtObj}.
\end{example}
 The latter approach, of applying a model-based optimizer to features extracted by a \ac{dnn} as in Example~\ref{exm:KLearnedFetaures}, can also be used to enforce decisions made by a \ac{dnn} to comply to some underlying physical requirements, see, e.g., \cite{zhao2021ensuring}. 


The above examples build upon the differentiability of the model-based solver to train the \ac{dnn} augmented into the method end-to-end. Nonetheless, \ac{dnn}-aided optimization can also augment model-based methods with \acp{dnn} that are pre-trained, possibly even in an unsupervised manner thus alleviating the dependence on the availability of labeled data. One such family of \ac{dnn}-aided optimization techniques, referred to as {\em plug-and-play networks} \cite{ahmad2020plug}, is exemplified next.
\begin{example}
\label{exm:PnPADMM}
Consider the application of \ac{admm} (Algorithm~\ref{alg:Algoadmm}) to solving \eqref{eqn:Recovery1}. Computing the proximal mapping in the second update step is often challenging, as the ability to evaluate the prior $\phi(\cdot)$ is required, which in practice may be unavailable or involve exhaustive computations. 
	Nonetheless, the proximal mapping is invariant of the task, and can be viewed as a denoiser for samples in $\mySet{S}$, e.g., high-resolution images for the setting in Example~\ref{exm:Inverse}. Denoisers are common \ac{dnn} models, which can be trained in an unsupervised manner, and can reliably operate on signals with intractable priors (e.g., natural images). By  letting $h_{\myVec{\theta}}(\cdot;\alpha)$ be a \ac{dnn} trained to denoise data in $\mySet{S}$ with noise level $\alpha$, one can thus implement Algorithm~\ref{alg:Algoadmm} without specifying the prior $\phi(\cdot)$ by replacing the proximal mapping with \cite{chan2016plug}
	\begin{equation}
	    \myVec{v}_{k+1}      =  h_{\myVec{\theta}}(\myVec{s}_{k+1} + \myVec{u}_k;\alpha_k).
	\end{equation}
	
\end{example}

The term {\em plug-and-play} is used to describe decision mappings as in Example~\ref{exm:PnPADMM} where pre-trained models are plugged into model-based optimizers without further tuning, as illustrated in Fig.~\ref{fig:ADMMComp}(c). Nonetheless, this methodology can also incorporate deep learning into the optimization procedure by, e.g., unfolding  the iterative optimization steps into a large \ac{dnn} whose trainable parameters are those of the smaller networks augmenting each iteration, as in \cite{gilton2019neumann}. This approach allows to benefit from both the ability of deep learning to implicitly represent complex domains, as well as the inference speed reduction  of deep unfolding along with its robustness to uncertainty and errors in the model parameters assumed to be known. Nonetheless, the fact that the iterative optimization must be learned from data in addition to the prior on $\mySet{S}$ implies that larger amounts of labeled data are required to train the system, compared to using the model-based optimizer.

An alternative approach to augment model-based solvers with pre-trained \acp{dnn} is the usage of {\em deep priors} \cite{bora2017compressed}. As opposed to plug-and-play networks, which augment the solver with a \ac{dnn} in order to cope with complex modelling, deep priors use \acp{dnn} to directly compute the (possibly intractable) decision rule objective, as shown in the next example.
\begin{example}
\label{exm:DeepPriors}
Consider again the setting in Example~\ref{exm:PnPADMM}, where one aims at solving \eqref{eqn:Recovery1} while the prior $\phi(\cdot)$ is unavailable and possibly intractable.  However, now let us assume that we have access to some bijective mapping from some latent space $\mySet{Z}$ to the signal space $\mySet{S}$, denoted $g:\mySet{Z}\mapsto \mySet{S}$, such that the prior term $\phi(\myVec{s})$ can be written in terms of $\myVec{z}$ as $ \phi(\myVec{s}) = \tilde{\phi}(\myVec{z})|_{\myVec{z} = G^{-1}(\myVec{s})}$. In this case, the \ac{map} rule in \eqref{eqn:Recovery1} becomes 
		\begin{align}
	    \Output 
	    &= g(\hat{\myVec{z}}), \quad  
	    \hat{\myVec{z}} =\mathop{\arg\min}\limits_{\myVec{z}} \frac{1}{2}\|\myVec{x}-\myMat{H}G(\myVec{z})\|^2_2 +\sigma^2\tilde{\phi}(\myVec{z}).
	    \label{eqn:RegOpt3alt}
	\end{align}
Deep generative priors \cite{bora2017compressed} use a pre-trained \ac{dnn}-based prior $h_{\myVec{\theta}}(\cdot)$, typically a generative network trained to map Gaussian vectors to $\mySet{S}$. The resulting objective becomes:	
\begin{equation}
      \hat{\myVec{z}} =\mathop{\arg\min}\limits_{\myVec{z}} \frac{1}{2}\|\myVec{x}-\myMat{H}h_{\myVec{\theta}}(\myVec{z})\|^2_2 + \lambda \|\myVec{z}\|^2_2 .
      \label{eqn:csLoss}
\end{equation}
Even though the exact formulation of $h_{\myVec{\theta}}(\cdot)$ may be highly complex, one can tackle \eqref{eqn:csLoss} via first-order optimization, building upon the fact that \acp{dnn} allow simple computation of gradients via backpropagation. These gradients are taken not with respect to the weights (as done in conventional \ac{dnn} training), but with respect to the input of the network.
\end{example}

{\bf Summary:} \ac{dnn}-aided optimizers implement decision boxes via an interleaving of model-based principled mathematical procedures and trained \acp{dnn}. The approach is particularly suitable for enabling decision making in complex environments with partial domain knowledge, where the latter is used to determine the suitable model-based optimizer, whose complex computations are replaced with \acp{dnn}. Such augmentations facilitate the model-based optimizer in coping with mismatches in its objective model and its parameters, and makes it applicable in complex domains.

Compared with the direct application of deep learning for the decision mappings, \ac{dnn}-aided optimizers are less generic and more task-specific due to the fact that they preserve the structure of a model-based optimizer. This property does not only facilitate their training procedure, which can sometimes be done unsupervised  as in Examples~\ref{exm:PnPADMM}-\ref{exm:DeepPriors}, but also yields  decision rules that are interpretable and suitable for their task. This interpretability can be exploited to extract additional measures of interest, e.g., uncertainty, as shown in \cite{klein2021uncertainty} for the \ac{dnn}-aided Kalman filter in Example~\ref{exm:KNet}; such measures, which are naturally obtained in model-based methods while being challenging to characterize for black-box \acp{dnn}, are often of importance in some applications.

\begin{figure*}
    \centering
    \includegraphics[width=0.85\linewidth]{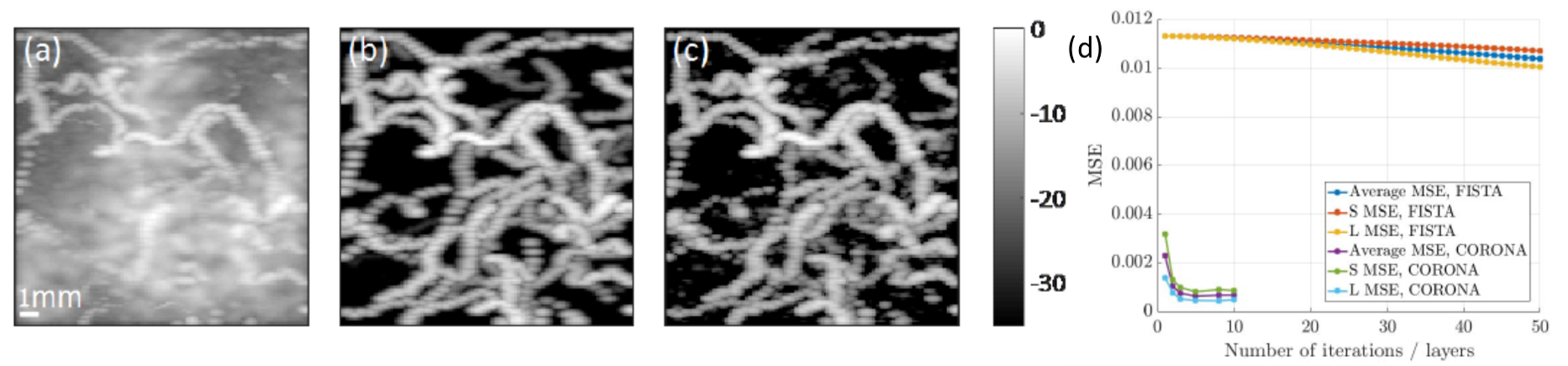}
    \caption{Experimental results (reproduced from \cite{solomon2019deep}) for recovering ultrasound contrast agents from cluttered maximum intensity projection images: $a)$ the observed image; $b)$ the ground-truth sparse contrast agents; $c)$ image recovered by deep unfolding; $d)$ \acs{mse} versus iterations/layers of deep unfolded network (CORONA) compared to  fast \ac{ista}.}
    \label{fig:DeepRPCA_Res2}
    \vspace{-0.2cm}
\end{figure*}

\section{Results}
In this section we experimentally exemplify  model-based deep learning methodology in a broad range of diverse application areas, including ultrasound imaging, optics, digital communications, and tracking of dynamic systems. 

\subsection{Ultrasound Imaging}
We first demonstrate the ability of deep unfolding, and particularly of \ac{lista}-like architectures as detailed in Example~\ref{exm:LISTA}, to facilitate the processing of ultrasound images. Our first example is taken from \cite{solomon2019deep}, which trained a deep unfolded decision box for clutter removal in contrast-enhanced ultrasound. Here the data was modeled as comprising a low-rank clutter background and a sparse blood flow image depicting the contrast agents. A generalization of \ac{ista} was then applied to robust principled component analysis (RPCA) optimization leading to an unfolded network referred to as CORONA: Convolutional rObust pRincipal cOmpoNent Analysis.
Here, both the context $\Input$ and the decision $\Output$ are maximum intensity projection ultrasound images; the decision rule type type is $K=10$ iterations of a generalized \ac{ista}, i.e., type~\ref{itm:iter}, whose objective parameters $\ObjParam$ and hyperparameters $\HypParam$ are tuned per-iteration from data via end-to-end training using the empirical risk \eqref{eqn:EmpRisk} with the $\ell_2$ loss $l_{\rm Est}(\cdot)$, computed over a  set of $\Ntraining = 4800$ images.

An experimental study of this application, showing that deep unfolding can infer both quickly and reliably, is presented in Fig.~\ref{fig:DeepRPCA_Res2}.  Fig.~\ref{fig:DeepRPCA_Res2}(c) shows the recovered  ultrasound (contrast agents) image from a cluttered image (Fig.~\ref{fig:DeepRPCA_Res2}(a)) achieved using deep unfolding of RPCA. Comparing the recovered image to the ground-truth in  Fig.~\ref{fig:DeepRPCA_Res2}(b) demonstrates the accuracy in using a \ac{dnn} to imitate the operations of the generalized \ac{ista} algorithm in a learned fashion. Furthermore, the fact that the unfolded network learns its parameters from data for each layer allows it to infer with a notably reduced number of layers compared to the corresponding number of iterations required by the model-based algorithm, which utilizes its full domain knowledge in applying the hard-coded iterative procedure. This is illustrated in Fig.~\ref{fig:DeepRPCA_Res2}(d) which demonstrates that the trained unfolded network can achieve with only a few layers a \ac{mse} accuracy which the model-based fast \ac{ista} of \cite{beck2009fast} does not approach even in $50$ iterations.

Deep unfolding can also be applied for super-resolution in ultrasound using micro bubbles. For instance, the work  \cite{bar2021learned} applied \ac{lista} for ultrasound-based breast lesion characterization. 
Here, the input $\Input$ is a low-resolution ultrasound image, while the decision $\Output$ is a high-resolution image. The decision rule is again an unfolded iterative algorithm, i.e., \ref{itm:iter}, with  $\ObjParam$ and $\HypParam$ jointly learned end-to-end as in Example~\ref{exm:LISTA}.

\begin{figure}
    \centering
    \includegraphics[width=0.8\columnwidth]{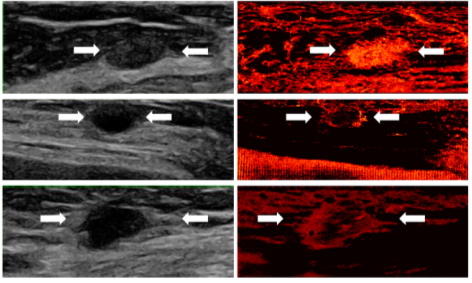}
    \caption{Experimental results (reproduced from \cite{bar2021learned}) for applying \ac{lista} for super-resolution in human scans of three lesions in breasts of three patients. Left: B-mode images; Right: super-resolution recoveries; Top: fibroadenoma (benign); Middle: cyst (benign); Bottom: invasive ductal carcinoma (malignant).}
    \label{fig:Ultrasound2}
    \vspace{-0.2cm}
\end{figure}

The ability of \ac{lista} to increase ultrasound resolution and facilitate diagnosis is demonstrated in Fig.~\ref{fig:Ultrasound2}. Here, a super-resolved recovery of a fibroadenoma (Fig.~\ref{fig:Ultrasound2}, top) shows an oval, well circumscribed mass with homogeneous high vascularization; a cyst (Fig.~\ref{fig:Ultrasound2}, middle) is visualized as a round structure with high concentration of blood vessels at the periphery of lesion; while an invasive ductal carcinoma (Fig.~\ref{fig:Ultrasound2}, bottom) shows an irregular mass with ill-defined margins, high concentration of blood vessels at the periphery of the mass, and a low concentrations of blood vessels at the center. These resolved features are not visually identifiable in the low-resolution.

\subsection{Microscopy Imaging}
Next, we demonstrate the application of model-based deep learning techniques in optics, considering again the usage of \ac{lista} (applied in the correlation domain)  for super-resolution. The context $\Input$ is a low resolution microscopy image, and $\Label$ is a high resolution image, with the decision rule being  $K=10$ iterations of \ac{ista} (\ref{itm:iter}) where the parameters $\ObjParam$ and $\HypParam$ are jointly learned from  data to minimize the empirical risk with the $\ell_2$ loss as its design objective. 

\begin{figure}
    \centering
    \includegraphics[width=\columnwidth]{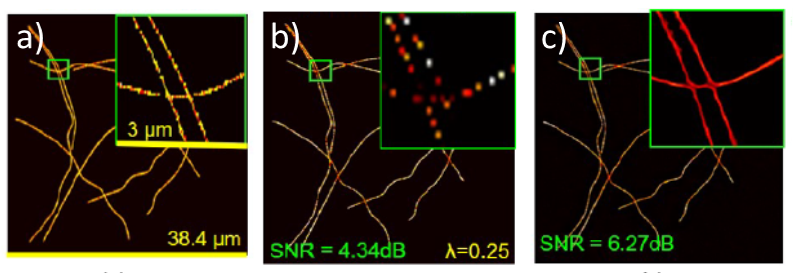}
    \caption{Sample results (reproduced from  \cite{sahel2021deep}) for applying deep unfolding for recovery of high resolution image. $a)$ simulated ground truth tubulin structure; $b)$ model-based  recovery with hyperparameter $\HypParam = 0.25$; $c)$ deep unfolded resolved image. }
    \label{fig:MicroImage1}
    \vspace{-0.2cm}
\end{figure}

Experimental results of applying the deep unfolded mapping trained for super-resolution in microscopy imaging are depicted in Fig.~\ref{fig:MicroImage1}, which is reproduced from  \cite{sahel2021deep} based on the method from \cite{dardikman2020learned}. Here, a super-resolved image is reconstructed from a simulated tubulins data set, composed of $350$ high-density frames, where the deep unfolded network (Fig.~\ref{fig:MicroImage1}(c)) is compared with $100$ iterations of the  model-based iterative sparse recovery algorithm from which it originates (Fig.~\ref{fig:MicroImage1}(b)). These results demonstrate the ability of deep unfolding, where both the objective parameters $\ObjParam$ and the hyperparameters $\HypParam$ are jointly learned end-to-end, to yield more abstract models that can overcome mismatches due to the  surrogate objectives of model-based optimization with complex data, where  mathematical descriptions are rarely accurate. 

\subsection{Digital Communications}
The experimental evaluations so far focused on deep unfolding methodology and on tasks where the context $\Input$ is an image. We proceed to a different family of tasks, arising in the operation of digital receivers, and present a  a numerical example for \ac{dnn}-aided optimization. We consider a scenario of symbol detection over causal stationary communication channels with finite memory, reproduced from \cite{shlezinger2020inference}. Here, the input $\Input$ is a real valued vector representing samples from an observed channel output, and $\Label$ is a vector of the transmitted symbols, whose entries take value in a discrete binary phase shift keying constellation.
The decision mapping which minimizes the error  is the \ac{map} rule which in such scenarios can be implemented with reduced complexity using the \ac{sp} algorithm \cite{loeliger2004introduction}.  This mapping relies on  accurate knowledge of the underlying channel which is captured using a factor graph.  The parameters of the decision rule are the weights of an internal \ac{dnn} used for evaluating the function nodes of the graph, and these parameters are tuned by minimizing the empirical cross entropy loss on a data set comprised of observations and their corresponding symbols.

 \begin{figure}
    \centering
    \includegraphics[width=0.9\columnwidth]{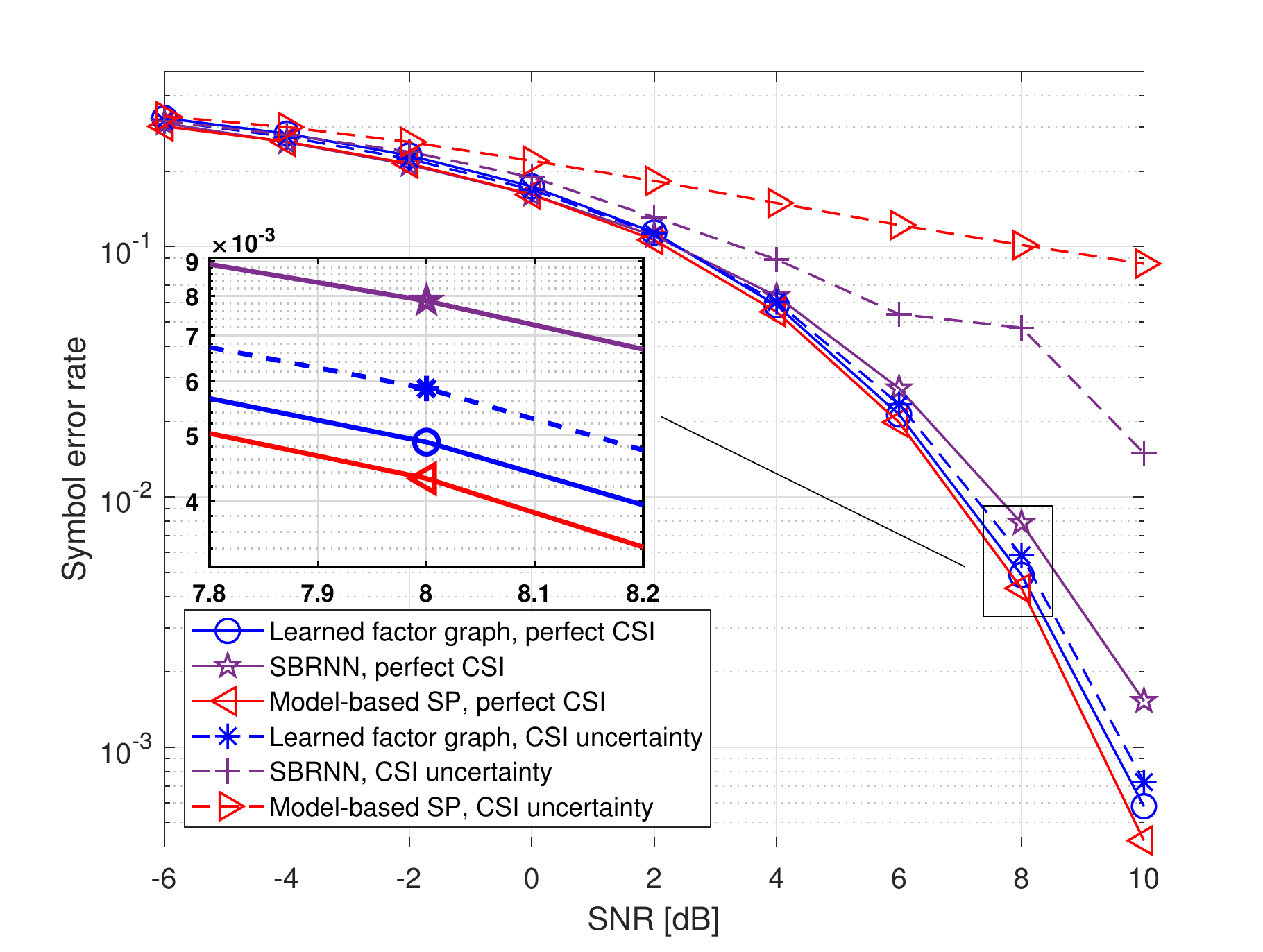}
    \caption{Experimental results from \cite{shlezinger2020inference} of learned factor graphs compared to the model-based \ac{sp} and the  sliding bidirectional \ac{rnn} (SBRNN) of \cite{farsad2018neural}. {\em Perfect CSI} implies that the system is trained and tested using
samples from the same channel, while in {\em CSI uncertainty} they are trained using samples from a set of different channels.}
    \label{fig:BCJRNet1}
    \vspace{-0.2cm}
\end{figure}

Fig.~\ref{fig:BCJRNet1} depicts the numerically evaluated \acl{ser} achieved by applying a \ac{dnn}-aided \ac{sp} algorithm where deep learning is used to learn to compute the function nodes of the factor graph from $\Ntraining=5000$ labeled samples. The results are compared to the performance of the model-based \ac{sp}, that requires complete knowledge of the underlying statistical model, as well as the sliding bidirectional \ac{rnn} detector proposed in \cite{farsad2018neural} for such setups, which utilizes a conventional \ac{dnn} architecture.  Fig.~\ref{fig:BCJRNet1} demonstrates the ability of learned factor graphs to enable accurate message passing inference in a data-driven manner, as the performance achieved using learned factor graphs approaches that of the \ac{sp} algorithm, which operates with full knowledge of the underlying statistical model. The numerical results also demonstrate that combining model-agnostic \acp{dnn} with model-aware optimization notably improves robustness to model uncertainty compared to applying the \ac{sp} algorithm with the inaccurate model. Furthermore, it also observed that the principled incorporation of \acp{dnn} and \ac{sp} inference allows to achieve improved performance compared to utilizing black-box \ac{dnn} architectures such as the sliding bidirectional \ac{rnn} detector, with limited training data.

\subsection{Tracking of Dynamic Systems} 
We conclude our experimental results with the application of \ac{dnn}-aided optimization for tracking of dynamic systems. 
Here, we use the \ac{dnn}-aided Kalman filter of Example~\ref{exm:KNet} to track the Lorentz attractor non-linear chaotic system. Both the context and the decision are three-dimensional vectors, representing  $3000$  noisy observations and the trajectory of the Lorenz attractor, respectively. The decision rule is a combination of a \ac{dnn} (\ref{itm:dnn}) and an affine mapping (\ref{itm:affine}) trained end-to-end from supervised data, as detailed in Example~\ref{exm:KNet}. We compare this model-based deep learning mapping with several model-based tracking algorithms designed for such settings -- the extended Kalman filter (EKF); unscented Kalman filter (UKF); and particle filter (PF) -- as well as to a black-box \ac{rnn} trained end-to-end.

\begin{table}
\begin{center}
\begin{tabular}{|c|c|c|c|c|c|c|}
\hline
&  EKF&  UKF &   PF & KalmanNet & \ac{rnn}\\
\hline
MSE [dB] &  -6.432 & -5.683 & -5.337 & {\bf -11.284} & 17.355\\
 \hline
 \textrm{Run-time} [sec] &  5.440
 & 6.072 & 62.946
 & {\bf 4.699} & 2.291\\
 \hline
\end{tabular}
\caption{\ac{mse} performance and run-time of the \ac{dnn}-aided KalmanNet, end-to-end \ac{rnn}, and the model-based EKF, UKF, and PF.}
\label{tbl:decimation}
\end{center}
\vspace{-0.4cm}
\end{table}

\begin{figure}
    \centering
    \includegraphics[width=0.9\columnwidth]{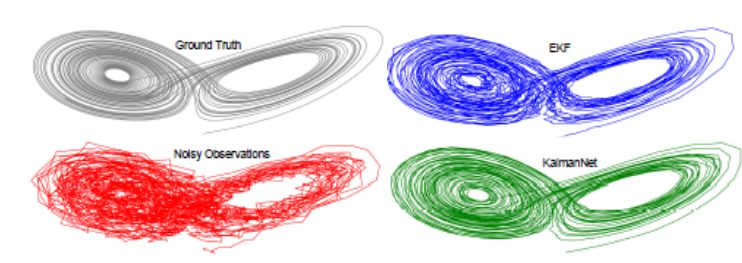}
    \caption{Tracking a single trajectory of the Lorentz attractor chaotic system using the \ac{dnn}-aided KalmanNet compared with the model-based EKF (reproduced from \cite{revach2021kalmannet}).}
    \label{fig:KNet}
\vspace{-0.2cm}
\end{figure}

 The results, reproduced from \cite{revach2021kalmannet} are summarized in Table~\ref{tbl:decimation}, and a representative reconstruction is visualized in Fig.~\ref{fig:KNet}.  It is observed in Table~\ref{tbl:decimation} that the gains of \ac{dnn}-aided optimization here are two-fold: first, it achieves the best \ac{mse} results due to its incorporation of the state-space model as domain knowledge along with a \ac{dnn} which learns to handle the complex dynamics and overcome the mismatches induced by the surrogate objective. Furthermore, the integration of deep learning allows \ac{dnn}-aided optimization to operate more quickly than its model-based counterparts, as some of the internal exhaustive computations of the algorithms are replaced with a \ac{dnn} inferring at fixed complexity.

\bibliographystyle{IEEEtran}
\bibliography{IEEEabrv,refs}

\begin{thebibliography}{10}
\providecommand{\url}[1]{#1}
\csname url@samestyle\endcsname
\providecommand{\newblock}{\relax}
\providecommand{\bibinfo}[2]{#2}
\providecommand{\BIBentrySTDinterwordspacing}{\spaceskip=0pt\relax}
\providecommand{\BIBentryALTinterwordstretchfactor}{4}
\providecommand{\BIBentryALTinterwordspacing}{\spaceskip=\fontdimen2\font plus
\BIBentryALTinterwordstretchfactor\fontdimen3\font minus
  \fontdimen4\font\relax}
\providecommand{\BIBforeignlanguage}[2]{{%
\expandafter\ifx\csname l@#1\endcsname\relax
\typeout{** WARNING: IEEEtran.bst: No hyphenation pattern has been}%
\typeout{** loaded for the language `#1'. Using the pattern for}%
\typeout{** the default language instead.}%
\else
\language=\csname l@#1\endcsname
\fi
#2}}
\providecommand{\BIBdecl}{\relax}
\BIBdecl

\bibitem{lecun2015deep}
Y.~LeCun, Y.~Bengio, and G.~Hinton, ``Deep learning,'' \emph{Nature}, vol. 521,
  no. 7553, p. 436, 2015.

\bibitem{Bengio09learning}
Y.~Bengio, ``Learning deep architectures for {AI},'' \emph{Foundations and
  trends in Machine Learning}, vol.~2, no.~1, pp. 1--127, 2009.

\bibitem{shlezinger2020model}
N.~Shlezinger, J.~Whang, Y.~C. Eldar, and A.~G. Dimakis, ``Model-based deep
  learning,'' arXiv preprint arXiv:2012.08405, 2020.

\bibitem{chen2021learning}
T.~Chen, X.~Chen, W.~Chen, H.~Heaton, J.~Liu, Z.~Wang, and W.~Yin, ``Learning
  to optimize: A primer and a benchmark,'' arXiv preprint arXiv:2103.12828,
  2021.

\bibitem{maier2022known}
A.~Maier, H.~K{\"o}stler, M.~Heisig, P.~Krauss, and S.~H. Yang, ``Known
  operator learning and hybrid machine learning in medical imaging—a review
  of the past, the present, and the future,'' \emph{Progress in Biomedical
  Engineering}, 2022.

\bibitem{monga2021algorithm}
V.~Monga, Y.~Li, and Y.~C. Eldar, ``Algorithm unrolling: Interpretable,
  efficient deep learning for signal and image processing,'' \emph{{IEEE}
  Signal Process. Mag.}, vol.~38, no.~2, pp. 18--44, 2021.

\bibitem{agrawal2021learning}
A.~Agrawal, S.~Barratt, and S.~Boyd, ``Learning convex optimization models,''
  \emph{{IEEE/CAA} J. Autom. Sinica}, vol.~8, no.~8, pp. 1355--1364, 2021.

\bibitem{ahmad2020plug}
R.~Ahmad, C.~A. Bouman, G.~T. Buzzard, S.~Chan, S.~Liu, E.~T. Reehorst, and
  P.~Schniter, ``Plug-and-play methods for magnetic resonance imaging: Using
  denoisers for image recovery,'' \emph{{IEEE} Signal Process. Mag.}, vol.~37,
  no.~1, pp. 105--116, 2020.

\bibitem{bora2017compressed}
A.~Bora, A.~Jalal, E.~Price, and A.~G. Dimakis, ``Compressed sensing using
  generative models,'' in \emph{International Conference on Machine
  Learning}.\hskip 1em plus 0.5em minus 0.4em\relax JMLR, 2017, pp. 537--546.

\bibitem{satorras2019combining}
V.~Garcia~Satorras, Z.~Akata, and M.~Welling, ``Combining generative and
  discriminative models for hybrid inference,'' \emph{Advances in Neural
  Information Processing Systems}, vol.~32, 2019.

\bibitem{shlezinger2021model}
N.~Shlezinger, N.~Farsad, Y.~C. Eldar, and A.~J. Goldsmith, ``Model-based
  machine learning for communications,'' arXiv preprint arXiv:2101.04726, 2021.

\bibitem{shalev2014understanding}
S.~Shalev-Shwartz and S.~Ben-David, \emph{Understanding machine learning: From
  theory to algorithms}.\hskip 1em plus 0.5em minus 0.4em\relax Cambridge
  university press, 2014.

\bibitem{boyd2004convex}
S.~P. Boyd and L.~Vandenberghe, \emph{Convex optimization}.\hskip 1em plus
  0.5em minus 0.4em\relax Cambridge university press, 2004.

\bibitem{boyd2011distributed}
S.~Boyd, N.~Parikh, and E.~Chu, \emph{Distributed optimization and statistical
  learning via the alternating direction method of multipliers}.\hskip 1em plus
  0.5em minus 0.4em\relax Now Publishers Inc, 2011.

\bibitem{parikh2014proximal}
N.~Parikh and S.~Boyd, ``Proximal algorithms,'' \emph{Foundations and Trends in
  optimization}, vol.~1, no.~3, pp. 127--239, 2014.

\bibitem{goodfellow2016deep}
I.~Goodfellow, Y.~Bengio, and A.~Courville, \emph{Deep learning}.\hskip 1em
  plus 0.5em minus 0.4em\relax MIT press, 2016.

\bibitem{vaswani2017attention}
A.~Vaswani, N.~Shazeer, N.~Parmar, J.~Uszkoreit, L.~Jones, A.~N. Gomez,
  {\L}.~Kaiser, and I.~Polosukhin, ``Attention is all you need,'' in
  \emph{Advances in Neural Information Processing Systems}, 2017, pp.
  5998--6008.

\bibitem{lecun1995convolutional}
Y.~LeCun and Y.~Bengio, ``Convolutional networks for images, speech, and time
  series,'' \emph{The handbook of brain theory and neural networks}, vol. 3361,
  no.~10, p. 1995, 1995.

\bibitem{mao2016image}
X.~Mao, C.~Shen, and Y.-B. Yang, ``Image restoration using very deep
  convolutional encoder-decoder networks with symmetric skip connections,''
  \emph{Advances in Neural Information Processing Systems}, vol.~29, 2016.

\bibitem{lillicrap2015continuous}
T.~P. Lillicrap, J.~J. Hunt, A.~Pritzel, N.~Heess, T.~Erez, Y.~Tassa,
  D.~Silver, and D.~Wierstra, ``Continuous control with deep reinforcement
  learning,'' arXiv preprint arXiv:1509.02971, 2015.

\bibitem{rumelhart1985learning}
D.~E. Rumelhart, G.~E. Hinton, and R.~J. Williams, ``Learning internal
  representations by error propagation,'' California Univ San Diego La Jolla
  Inst for Cognitive Science, Tech. Rep., 1985.

\bibitem{agrawal2020learning}
A.~Agrawal, S.~Barratt, S.~Boyd, and B.~Stellato, ``Learning convex
  optimization control policies,'' in \emph{Learning for Dynamics and
  Control}.\hskip 1em plus 0.5em minus 0.4em\relax PMLR, 2020, pp. 361--373.

\bibitem{agrawal2019differentiable}
A.~Agrawal, B.~Amos, S.~Barratt, S.~Boyd, S.~Diamond, and J.~Z. Kolter,
  ``Differentiable convex optimization layers,'' \emph{Advances in Neural
  Information Processing Systems}, vol.~32, 2019.

\bibitem{maclaurin2015gradient}
D.~Maclaurin, D.~Duvenaud, and R.~Adams, ``Gradient-based hyperparameter
  optimization through reversible learning,'' in \emph{International Conference
  on Machine Learning}, 2015, pp. 2113--2122.

\bibitem{lorraine2020optimizing}
J.~Lorraine, P.~Vicol, and D.~Duvenaud, ``Optimizing millions of
  hyperparameters by implicit differentiation,'' in \emph{International
  Conference on Artificial Intelligence and Statistics}.\hskip 1em plus 0.5em
  minus 0.4em\relax PMLR, 2020, pp. 1540--1552.

\bibitem{barratt2020fitting}
S.~T. Barratt and S.~P. Boyd, ``Fitting a {K}alman smoother to data,'' in
  \emph{2020 American Control Conference (ACC)}.\hskip 1em plus 0.5em minus
  0.4em\relax IEEE, 2020, pp. 1526--1531.

\bibitem{sutskever2013training}
I.~Sutskever, \emph{Training recurrent neural networks}.\hskip 1em plus 0.5em
  minus 0.4em\relax University of Toronto, 2013.

\bibitem{xu2021ekfnet}
L.~Xu and R.~Niu, ``{EKFN}et: Learning system noise statistics from measurement
  data,'' in \emph{IEEE International Conference on Acoustics, Speech and
  Signal Processing}, 2021, pp. 4560--4564.

\bibitem{gregor2010learning}
K.~Gregor and Y.~LeCun, ``Learning fast approximations of sparse coding,'' in
  \emph{International Conference on International Conference on Machine
  Learning}, 2010, pp. 399--406.

\bibitem{johnston2021admm}
J.~Johnston, Y.~Li, M.~Lops, and X.~Wang, ``{ADMM-N}et for communication
  interference removal in stepped-frequency radar,'' \emph{{IEEE} Trans. Signal
  Process.}, vol.~69, pp. 2818--2832, 2021.

\bibitem{khobahi2021model}
S.~Khobahi, N.~Shlezinger, M.~Soltanalian, and Y.~C. Eldar, ``{LoRD-Net}: Low
  resolution detection network for deep low-resolution receivers,''
  \emph{{IEEE} Trans. Signal Process.}, vol.~69, pp. 5651--5664, 2021.

\bibitem{revach2021kalmannet}
G.~Revach, N.~Shlezinger, X.~Ni, A.~L. Escoriza, R.~J. van Sloun, and Y.~C.
  Eldar, ``Kalman{N}et: Neural network aided {K}alman filtering for partially
  known dynamics,'' \emph{{IEEE} Trans. Signal Process.}, vol.~70, pp.
  1532--1547, 2022.

\bibitem{coskun2017long}
H.~Coskun, F.~Achilles, R.~DiPietro, N.~Navab, and F.~Tombari, ``Long
  short-term memory {Kalman} filters: Recurrent neural estimators for pose
  regularization,'' in \emph{Proceedings of the IEEE International Conference
  on Computer Vision}, 2017, pp. 5524--5532.

\bibitem{zhao2021ensuring}
T.~Zhao, X.~Pan, M.~Chen, and S.~H. Low, ``Ensuring {DNN} solution feasibility
  for optimization problems with convex constraints and its application to {DC}
  optimal power flow problems,'' arXiv preprint arXiv:2112.08091, 2021.

\bibitem{chan2016plug}
S.~H. Chan, X.~Wang, and O.~A. Elgendy, ``Plug-and-play {ADMM} for image
  restoration: Fixed-point convergence and applications,'' \emph{IEEE
  Transactions on Computational Imaging}, vol.~3, no.~1, pp. 84--98, 2016.

\bibitem{gilton2019neumann}
D.~Gilton, G.~Ongie, and R.~Willett, ``Neumann networks for inverse problems in
  imaging,'' \emph{IEEE Transactions on Computational Imaging}, vol.~6, pp.
  328--343, 2019.

\bibitem{klein2021uncertainty}
I.~Klein, G.~Revach, N.~Shlezinger, J.~E. Mehr, R.~J. van Sloun, and Y.~Eldar,
  ``Uncertainty in data-driven {K}alman filtering for partially known
  state-space models,'' in \emph{IEEE International Conference on Acoustics,
  Speech and Signal Processing}, 2022.

\bibitem{solomon2019deep}
O.~Solomon, R.~Cohen, Y.~Zhang, Y.~Yang, Q.~He, J.~Luo, R.~J. van Sloun, and
  Y.~C. Eldar, ``Deep unfolded robust {PCA} with application to clutter
  suppression in ultrasound,'' \emph{{IEEE} Trans. Med. Imag.}, vol.~39, no.~4,
  pp. 1051--1063, 2019.

\bibitem{beck2009fast}
A.~Beck and M.~Teboulle, ``A fast iterative shrinkage-thresholding algorithm
  for linear inverse problems,'' \emph{SIAM journal on imaging sciences},
  vol.~2, no.~1, pp. 183--202, 2009.

\bibitem{bar2021learned}
O.~Bar-Shira, A.~Grubstein, Y.~Rapson, D.~Suhami, E.~Atar, K.~Peri-Hanania,
  R.~Rosen, and Y.~C. Eldar, ``Learned super resolution ultrasound for improved
  breast lesion characterization,'' in \emph{International Conference on
  Medical Image Computing and Computer-Assisted Intervention}.\hskip 1em plus
  0.5em minus 0.4em\relax Springer, 2021, pp. 109--118.

\bibitem{sahel2021deep}
Y.~B. Sahel, J.~P. Bryan, B.~Cleary, S.~L. Farhi, and Y.~C. Eldar, ``Deep
  unrolled recovery in sparse biological imaging,'' \emph{{IEEE} Signal
  Process. Mag.}, vol.~39, no.~2, pp. 45--57, 2022.

\bibitem{dardikman2020learned}
G.~Dardikman-Yoffe and Y.~C. Eldar, ``Learned {SPARCOM}: unfolded deep
  super-resolution microscopy,'' \emph{Optics express}, vol.~28, no.~19, pp.
  27\,736--27\,763, 2020.

\bibitem{shlezinger2020inference}
N.~Shlezinger, N.~Farsad, Y.~C. Eldar, and A.~J. Goldsmith, ``Learned factor
  graphs for inference from stationary time sequences,'' \emph{{IEEE} Trans.
  Signal Process.}, vol.~70, pp. 366--380, 2021.

\bibitem{loeliger2004introduction}
H.-A. Loeliger, ``An introduction to factor graphs,'' \emph{{IEEE} Signal
  Process. Mag.}, vol.~21, no.~1, pp. 28--41, 2004.

\bibitem{farsad2018neural}
N.~Farsad and A.~Goldsmith, ``Neural network detection of data sequences in
  communication systems,'' \emph{{IEEE} Trans. Signal Process.}, vol.~66,
  no.~21, pp. 5663--5678, 2018.

\end{thebibliography}

\begin{IEEEbiography}[{\includegraphics[width=1in,height=1.25in,clip,keepaspectratio]{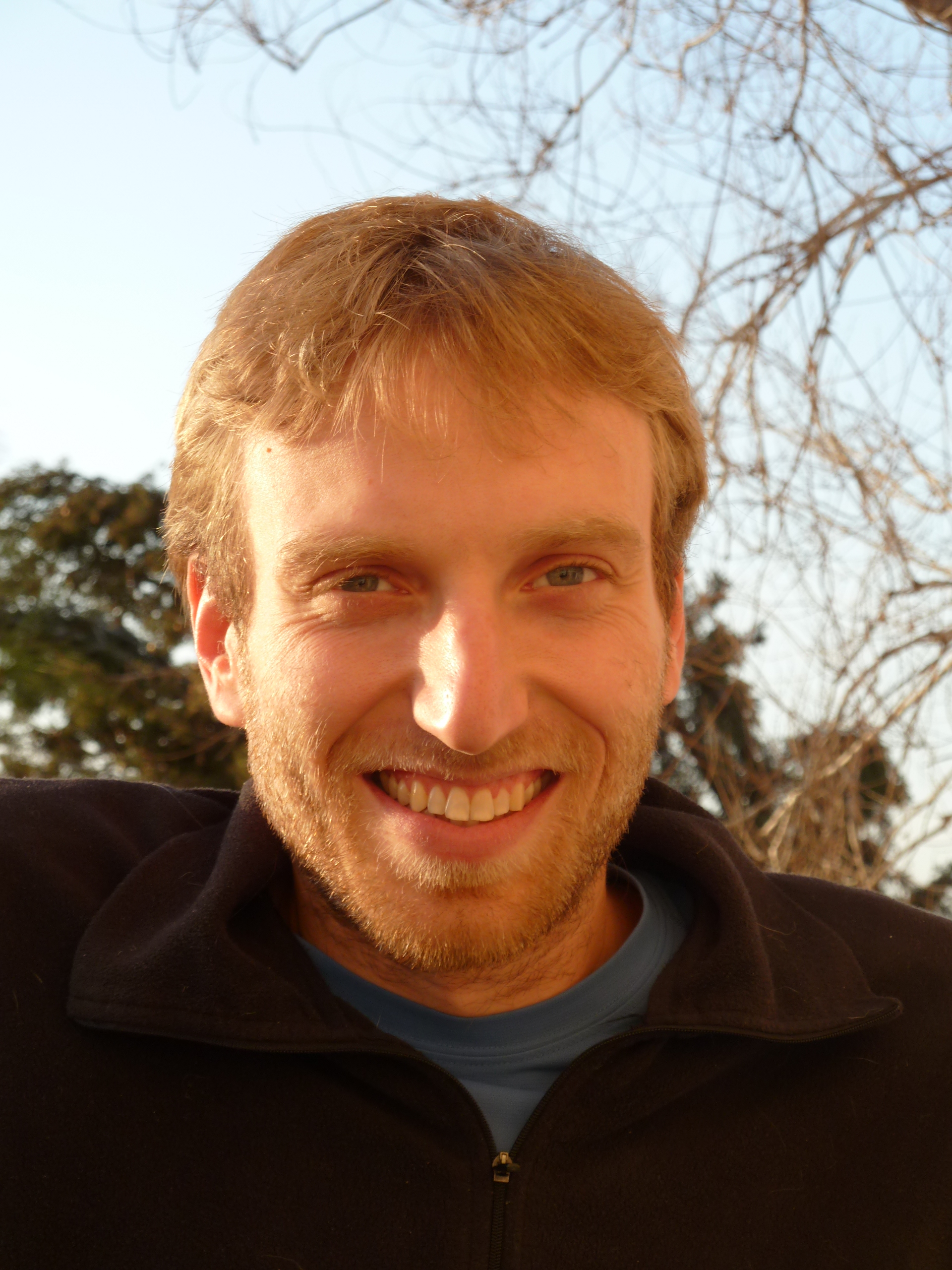}}]{Nir Shlezinger} (M’17) is an assistant professor in the School of Electrical and Computer Engineering in Ben-Gurion University, Israel. He received his B.Sc., M.Sc., and Ph.D. degrees in 2011, 2013, and 2017, respectively, from Ben-Gurion University, Israel, all in electrical and computer engineering.
From 2017 to 2019 he was a postdoctoral researcher in the Technion, and from 2019 to 2020 he was a postdoctoral researcher in Weizmann Institute of Science, where he was awarded the FGS prize for outstanding research achievements.
His research interests include communications, information theory, signal processing, and machine learning. 
\end{IEEEbiography}

\begin{IEEEbiography}[{\includegraphics[width=1in,height=1.25in,clip,keepaspectratio]{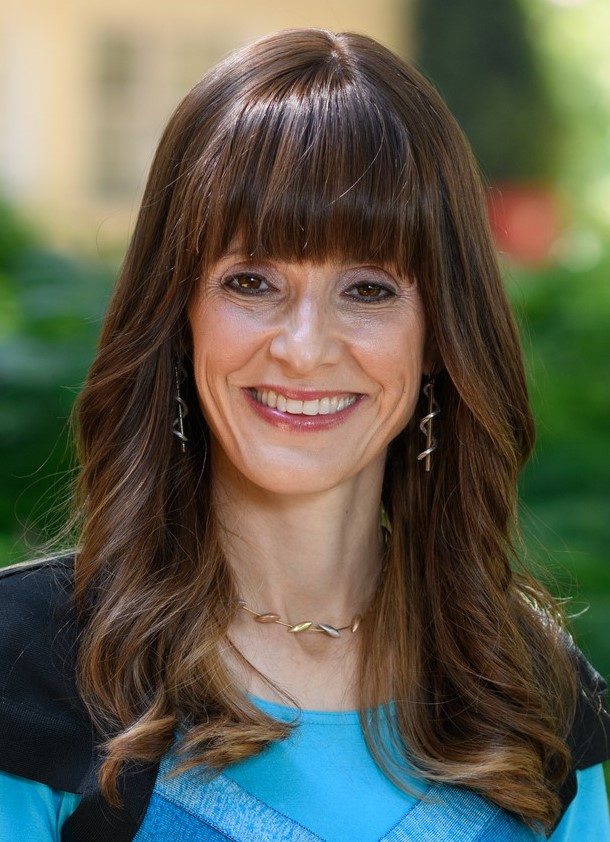}}]{Yonina C. Eldar} (S’98–M’02–SM’07–F’12) received the B.Sc. degree in Physics in 1995 and the B.Sc. degree in Electrical Engineering in 1996 both from Tel-Aviv University (TAU), Tel-Aviv, Israel, and the Ph.D. degree in Electrical Engineering and Computer Science in 2002 from the Massachusetts Institute of Technology (MIT), Cambridge.
She is currently a Professor in the Department of Mathematics and Computer Science, Weizmann Institute of Science, Rehovot, Israel. She was previously a Professor in the Department of Electrical Engineering at the Technion, where she held the Edwards Chair in Engineering. She is also a Visiting Professor at MIT, a Visiting Scientist at the Broad Institute, and an Adjunct Professor at Duke University and was a Visiting Professor at Stanford. She is a member of the Israel Academy of Sciences and Humanities (elected 2017), an IEEE Fellow and a EURASIP Fellow. Her research interests are in the broad areas of statistical signal processing, sampling theory and compressed sensing, learning and optimization methods, and their applications to biology, medical imaging and optics.

Dr. Eldar has received many awards for excellence in research and teaching, including the  IEEE Signal Processing Society Technical Achievement Award (2013), the IEEE/AESS Fred Nathanson Memorial Radar Award (2014), and the IEEE Kiyo Tomiyasu Award (2016). She was a Horev Fellow of the Leaders in Science and Technology program at the Technion and an Alon Fellow. She received the Michael Bruno Memorial Award from the Rothschild Foundation, the Weizmann Prize for Exact Sciences, the Wolf Foundation Krill Prize for Excellence in Scientific Research, the Henry Taub Prize for Excellence in Research (twice), the Hershel Rich Innovation Award (three times), the Award for Women with Distinguished Contributions, the Andre and Bella Meyer Lectureship, the Career Development Chair at the Technion, the Muriel \& David Jacknow Award for Excellence in Teaching, and the Technion’s Award for Excellence in Teaching (two times).  She received several best paper awards and best demo awards together with her research students and colleagues including the SIAM outstanding Paper Prize, the UFFC Outstanding Paper Award, the Signal Processing Society Best Paper Award and the IET Circuits, Devices and Systems Premium Award, was selected as one of the 50 most influential women in Israel and in Asia, and is a highly cited researcher.
She was a member of the Young Israel Academy of Science and Humanities and the Israel Committee for Higher Education. She is the Editor in Chief of Foundations and Trends in Signal Processing, a member of the IEEE Sensor Array and Multichannel Technical Committee and serves on several other IEEE committees. In the past, she was a Signal Processing Society Distinguished Lecturer, member of the IEEE Signal Processing Theory and Methods and Bio Imaging Signal Processing technical committees, and served as an associate editor for the IEEE Transactions On Signal Processing, the EURASIP Journal of Signal Processing, the SIAM Journal on Matrix Analysis and Applications, and the SIAM Journal on Imaging Sciences. She was Co-Chair and Technical Co-Chair of several international conferences and workshops. She is author of the book "Sampling Theory: Beyond Bandlimited Systems" and co-author of five other books published by Cambridge University Press.
\end{IEEEbiography}

\begin{IEEEbiography}[{\includegraphics[width=1in,height=1.25in,clip,keepaspectratio]{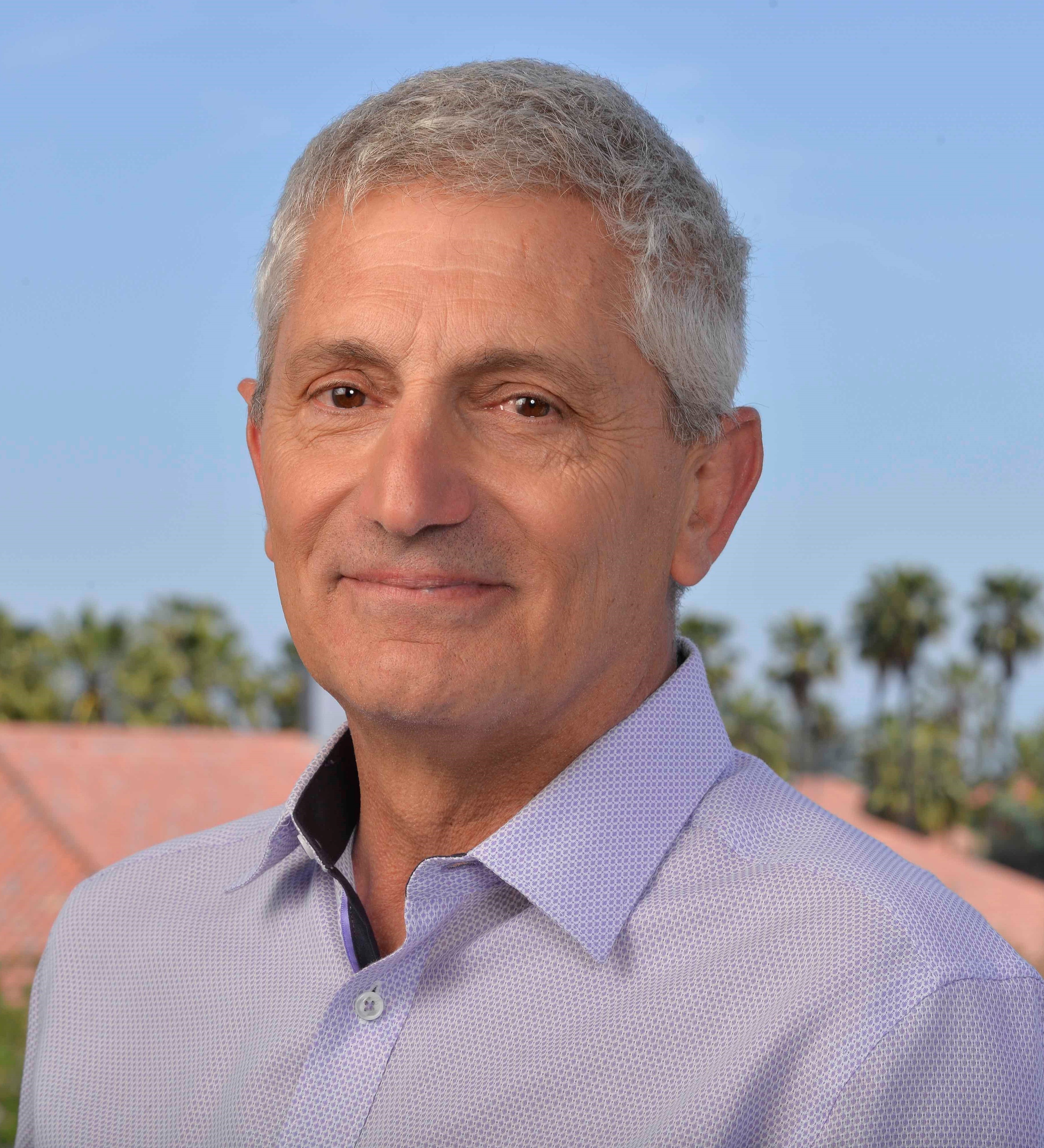}}]{Stephen P. Boyd} (F’ 99) is the Samsung Professor of
engineering, and Professor of electrical engineering
in the Information Systems Laboratory at Stanford
University, with courtesy appointments in computer
science and management science and engineering.
He received the A.B. degree in mathematics from
Harvard University, USA, in 1980, and the Ph.D. in
electrical engineering and computer science from the
University of California, USA, in 1985, and then
joined the faculty at Stanford. His current research
interests include convex optimization applications in control, signal
processing, machine learning, finance, and circuit design.
\end{IEEEbiography}

\EOD

\end{document}